\documentclass[12pt,a4paper]{article}
\usepackage[T1]{fontenc}
\usepackage[latin9]{inputenc}
\usepackage{xspace}
\usepackage{amsmath}
\usepackage{amssymb}
\usepackage{epsfig}
\usepackage{graphicx}
\usepackage{psfrag,pstricks}
\usepackage[left=2cm,right=2cm,top=2.5cm,bottom=2.5cm]{geometry}

\def\lsim{\raise0.3ex\hbox{$\;<$\kern-0.75em\raise-1.1ex\hbox{$\sim\;$}}}
\def\gsim{\raise0.3ex\hbox{$\;>$\kern-0.75em\raise-1.1ex\hbox{$\sim\;$}}}

\DeclareMathAlphabet   {\mathsc}{OT1}{cmr}{m}{sc}

\def\[{\left [}
\def\]{\right ]}
\def\({\left (}
\def\){\right )}

\newcommand{\gappeq}{\mathrel{\rlap {\raise.5ex\hbox{$>$}}
{\lower.5ex\hbox{$\sim$}}}}
\newcommand{\lappeq}{\mathrel{\rlap{\raise.5ex\hbox{$<$}}
{\lower.5ex\hbox{$\sim$}}}}

\newcommand{\bea}{\begin{eqnarray}}
\newcommand{\eea}{\end{eqnarray}}

\newrgbcolor{darkred}{0.7 0 0}
\newrgbcolor{green}{0 1 0}
\newrgbcolor{darkblue}{0 0 0.7}
 \newrgbcolor{purple}{1 0 1}

\begin{document}

\vspace{-1truecm}

\rightline{LPT--Orsay 08/38}
\rightline{FTUAM 08/6}
\rightline{IFT-UAM/CSIC-08-21}
\rightline{IFT-6/2008}

\vspace{0.cm}

\begin{center}

{\Large {\bf Determining the WIMP mass using the complementarity between direct and
indirect searches and the ILC}}
\vspace{0.3cm}\\

{\large N. Bernal$^1$, A. Goudelis$^1$,Y. Mambrini$^2$, C. Mu\~noz$^{3,4}$
}
\vspace{0.3cm}\\

$^1$
Laboratoire de Physique Th\'eorique,
Universit\'e Paris-Sud, F-91405 Orsay, France
\vspace{0.3cm}\\

$^2$
Institute of Theoretical Physics, Warsaw University,
ul. Hoza 69, 00-681 Warsaw, Poland
\vspace{0.3cm}\\

$^3$
Departamento de F\'isica Te\'orica C-XI,
Universidad Aut\'onoma de Madrid,\\
%and Instituto de F\'isica
%Te\'orica C--XVI
%,
%\\
Cantoblanco, 28049 Madrid, Spain
\vspace{0.3cm}\\

$^4$
%Departamento de F\'isica Te\'orica C-XI and
Instituto de F\'isica Te\'orica UAM/CSIC,
Universidad Aut\'onoma de Madrid,\\
Cantoblanco,
28049 Madrid, Spain
\vspace{0.3cm}\\

\end{center}

\vspace{-0.5cm}

\abstract{We study the possibility of identifying dark matter properties from
XENON--like $100$ kg experiments and the GLAST satellite mission. We show that
whereas direct detection experiments will probe efficiently
light WIMPs, given a positive detection (at the 10\% level for
$m_{\chi} \lesssim 50$ GeV), GLAST will be able to confirm
and even increase the precision in the case of a NFW profile, for a
WIMP-nucleon cross-section $\sigma_{\chi-p} \lesssim 10^{-8}$ pb.
We also predict the production rate of a WIMP in the next generation
of colliders (ILC), and compare their sensitivity to the WIMP mass
with the XENON and GLAST projects.

}

\newpage

\vspace{3cm}

\newpage

\tableofcontents

\newpage

\section{Introduction}

%{\bf YANN : three papers you have to read with this draft :
%A.M. Green hep-ph/0703217, Hooper : 0711.4621, Drees : 0710.4296
%This papers are very recent, especially the two last ones, and
%appears during the completion of the work. However, it is important
%to underline that we recover the results of the 3 papers. Not
%one of them address the problem of the complementarity of both direct
%and indirect detection modes.
%}

There exists strong evidence that a large
fraction of the matter in our Universe
is non-luminous \cite{mireview}.
Such evidence includes the motion of cluster member galaxies
\cite{evidence1},
gravitational lensing \cite{evidence2},
cosmic microwave background \cite{evidence3}, observations
of the flat rotation curves of galaxies \cite{evidence4}, etc.
Dark matter plays a central role in current structure formation
theories, and its microscopic properties have significant impact
on the spatial distribution of mass, galaxies and clusters.
Unraveling the nature of dark matter is therefore of critical importance.
A Weakly Interacting Massive Particle (WIMP), with mass lying
from the GeV to the TeV scale, is one of the preferred candidates for the dark matter of the Universe.

Different experimental programs are developing huge efforts
to observe and identify the particle nature of dark matter.
This can be achieved by direct measurement of the recoil energy
of a nucleus when scattered by a WIMP, or indirectly via the observation of WIMP
annihilation products.
In both cases, the sensitivity depends strongly on the
background and on the theoretical assumptions of the model.
It would be interesting to combine all these efforts
to invent intelligent strategies for determining
the nature of dark matter \cite{Hooper:2008sn}.
Recently, several works (see, for example,  Refs. \cite{Green:2007rb} and \cite{Green:2008rd}
for the case of direct detection and \cite{Baltz:2008wd, Dodelson:2007gd}
for the indirect detection case) have shown that precision
measurements of the mass of WIMPs are not only reserved to
the domain of accelerator physics. In all of these studies, model
independent bounds are derived for annihilation cross-sections, masses
or WIMP--nucleus scattering cross-sections.
The drawback of a model-independent framework
(lack of determined microscopic processes) is largely compensated by the
 universality of the method: instead of restricting a theoretical parameter
space, we restrict observable physical quantities (masses, branching ratios).
Indeed, these limits are valid for a great number of WIMP candidates such as, for example, the
supersymmetric neutralino, the lightest Kaluza Klein excitation, etc.

The aim of the present work is to analyse two of the most promising
experiments, XENON \cite{Angle:2007uj} and GLAST \cite{GLAST}, calculating and
comparing their sensitivity to a WIMP mass depending on the astrophysical
hypothesis (velocity distribution of WIMPs, density profile of the galactic halo).
In addition, using the known cosmological abundance of dark matter in the Universe,
we estimate the radiative WIMP production rate in the next generation
of colliders (ILC) and compare their sensitivity to the WIMP mass
with the XENON and GLAST projects.

Our goal is not to perform an exhaustive analysis of the mass determination
capacity of each detection mode. The GLAST discovery potential has been studied in
much detail in ref.\cite{Baltz:2008wd}  whereas direct detection experiments
are thoroughly treated in refs. \cite{Green:2007rb} and \cite{Green:2008rd}. Instead,
we are mostly interested in examining at which point and under which circumstances
the different detection techniques can achieve comparable precision, thus acting in
a complementary way (an issue which has been commented upon in a great number of references
but has never been treated as such).
In the case of indirect
detection, we will restrict ourself to $W^+ W^-$ WIMP annihilation final states.
It is well known, and is demonstrated in the Appendix, that different
Standard Model particles contribute differently in the gamma-ray spectrum,
the most prominent example being leptonic final states which tend to produce
harder gamma-ray spectrae. Nevertheless, we should say in advance that
at least in the ILC section we include an electron-positron final state,
whereas the $W^+ W^-$ spectrum can be considered quite representative of
bosonic and hadronic ones.

The paper is organised as follows.
In section 2 we discuss the event rate and WIMP-nucleon scattering cross-section
for a XENON-like experiment, in a microscopically model-independent approach. In Section 3 we
carry out a similar analysis for the GLAST experiment, discussing in this case the WIMP annihilation cross
section, and taking into account different halo profiles.
Section 4 is dedicated to the comparison between these two modes of detection. In Section 5 we analyse the
sensitivity that we can expect in such a
model-independent framework for a linear collider.
Finally, in Section 6 we carry out the comparison between the three detection modes.
The conclusions are left for Section 7.

\section{Direct detection}
\subsection{Differential event rate}
\label{XENON}

In spite of the experimental challenges, a number of efforts worldwide
are actively pursuing to directly detect WIMPs with a variety of targets
and approaches. Many direct dark matter detection experiments are
now either operating or in preparation.
% In these experiments, numerous approaches have been used to measure
% efficiently the recoil energy from a WIMP-nucleus scattering, from
% the observation of photons (CUORICINO \cite{Arnaboldi:2003tu}), scintillation
% (DAMA \cite{Bernabei:2000qi},
%  ZEPLIN \cite{Sumner:2007zz})
% or ionization (HDMS \cite{KlapdorKleingrothaus:2002pg}).
% Another generation of detectors rely on
% more powerful discrimination methods, using various schemes to extract
% as much information as possible from the target--detector.
% Cryogenic detectors are based on the simultaneous measurement of
% ionization and phonons (CDMS \cite{ArmelFunkhouser:2005zy},
% EDELWEISS \cite{Sanglard:2005we}),
% phonons and scintillation (CRESST \cite{Bravin:1999fc})
% or ionization and scintillation (XENON \cite{Angle:2007uj}).
All these experiments measure the number $N$ of elastic
collisions between WIMPs and target nuclei in a detector,
per unit detector mass and per unit of time, as a function of the
nuclear recoil energy $E_r$.
The detection rate in a detector depends on the density
$\rho_0\simeq0.3$ GeV cm$^{-3}$ and velocity distribution $f(v_\chi)$ of WIMPs near the Earth.
In general, the differential event rate per unit detector mass and
per unit of time can be written as:
\begin{equation}
\frac{dN}{dE_r}=\frac{\sigma_{\chi-N}\,\rho_0}{2\,m_r^2\,m_\chi}\,
F(E_r)^2\int_{v_{min}(E_r)}^{\infty}\frac{f(v_\chi)}{v_\chi}dv_\chi\ ,
\label{Recoil}
\end{equation}

\noindent
where the WIMP-nucleus cross section, $\sigma_{\chi-N}$, is related to the
WIMP-nucleon cross section, $\sigma_{\chi-p}$, by $\sigma_{\chi-N} = \sigma_{\chi-p}
(A m_r / M_r)^2$, with $M_r =\frac{m_\chi  m_p}{m_\chi + m_p}$ the WIMP-nucleon reduced mass, $m_r=\frac{m_\chi\,m_N}{m_\chi+m_N}$ the
WIMP-nucleus reduced mass, $m_\chi$ the WIMP mass, $m_N$ the nucleus mass,
and $A$ the atomic weight. $F$ is the form factor.

For the velocity distribution we take a simple Maxwellian halo
\begin{equation}
f(v_\chi)\,d^3v_\chi=\frac{1}{(v_{\chi}^0)^3\pi^{3/2}}\,e^{-(v_\chi/v_{\chi}^0)^2}\,d^3v_\chi\ ,
\end{equation}

\noindent
where $v_{\chi}^0 = 220\pm 20$ km/s is the velocity of the Sun around the
galactic center with its uncertainty, and we have neglected the
motion of the Earth around the Sun. After integrating over the angular
part in order to find the speed distribution we get:

\begin{equation}
f(v_\chi)\,dv_\chi=\frac{4\,v_\chi^2}{(v_{\chi}^0)^3\sqrt{\pi}}\,
e^{-(v_\chi/v_{\chi}^0)^2}\,dv_\chi\ ,
\end{equation}

\noindent
The integration over velocities is limited to those which
can give place to a recoil energy $E_r$, thus there is a minimal velocity
given by $v_{min}(E_r)=\sqrt{\frac{m_N\,E_r}{2\,m_r^2}}$.

\begin{figure}[!]
    \begin{center}
\centerline{
%\vskip -4.truecm
       \epsfig{file=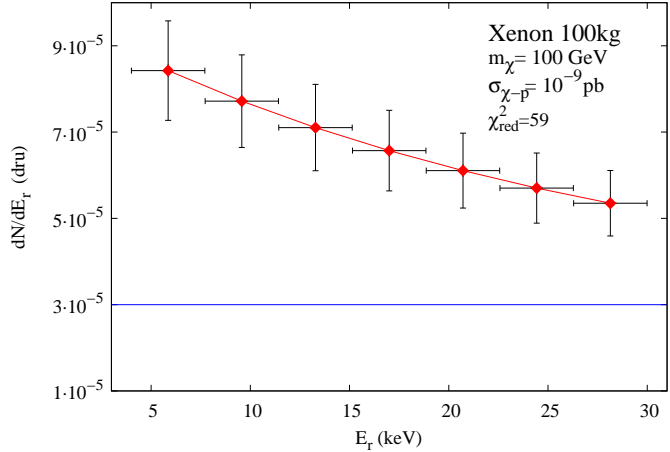,angle=-90,width=0.5\textwidth}
       }
          \caption{{\footnotesize
          XENON event rate expectations for the case of
          Differential rate versus recoil energy of the nucleus for a
 WIMP mass $m_{\chi}=100$ GeV and cross section $\sigma_{\chi-p} = 10^{-9}$ pb.
The error bars shown are those expected for the XENON $100$ kg experiment
after $3$ years of observation. The lower (blue) line is the
background--only prediction.
The $\chi^2$ per degree of freedom ($\chi^2_{\mathrm{red}}$) is 59, giving
a signal clearly distinguishable from the background.
}}
        \label{fig:XENON}
    \end{center}

\end{figure}

The effective interaction between the WIMP and a nucleus is given by the
Woods-Saxon form factor

\begin{equation}
F(E_r)=\frac{3\,j_1(q\,R_1)}{q\,R_1}\,e^{-(q\,s)^2}\ ,
\end{equation}
where the transferred momentum is $q=\sqrt{2\,m_N\,E_r}$, $j_1$ is a spherical, first-order Bessel function,
 $R_1=\sqrt{R^2-5\,s^2}$ with $R\simeq1.2\cdot A^{1/3}$ fm, $A$ is
the mass number, and $s\simeq 1$ fm.

In Fig.\ref{fig:XENON} we show an example of a signal
with a standard neutron background in a XENON--like ($100$ kg)
experiment, after $3$ years of data acquisition, as a function of the recoil
energy.
For a WIMP mass of $100$ GeV
and a WIMP--nucleon cross-section of $10^{-9}$ pb, such an experiment
would reach a pretty large $\chi^2$ per degree of freedom ($\chi^2_{\mathrm{red}}$) , of the order of 60.

\subsection{The statistical method}
\label{statistical}
In order to plot discrimination capacity regions, we use a method inspired by the treatment presented in
\cite{Green:2007rb}.
For a WIMP with mass $m_{\chi}^{real}$ and a WIMP-proton scattering cross-section $\sigma_{\chi-p}^{real}$
(in the case of indirect detection $\sigma_{\chi-p}$ will, of course, be replaced by the total thermally
averaged annihilation cross-section $\left\langle \sigma v \right\rangle$), we can calculate the theoretically
expected number of events, say $N_{th}$, simply by integrating Eq.(\ref{Recoil}) from $E_{th}$, the threshold energy
which we consider as $4$ keV, up to $E_{sup}$, the maximal observable energy, which we take as equal to $30$ keV.
Now, in a real-life experiment, one would expect the observed number of events to slightly deviate from this
ideal value giving, say, $N_{Exp}$ events. In order to approach a more realistic situation, we perform our estimations
not starting from $N_{th}$, but rather from $N_{Exp}$, where the ``experimental'' number of events is drawn from a
Poisson distribution with mean value $N_{th}$. Then, we can Monte-Carlo generate  $N_{Exp}$ events according to Eq.(\ref{Recoil})
and we obtain what could actually resemble to an experimental spectrum. Such a set of events, together with the corresponding value of
$N_{Exp}$ will be in the following referred to as an ``experiment''.

Then, for every point in the parameter space $(m_{\chi}, \sigma_{\chi-p})$, we calculate the corresponding extended likelihood
function which is given by:
\begin{equation}
L = \frac{(N_{th}^{scan})^{N_{Exp}}}{N_{Exp}!}\exp{(-N_{th}^{scan})}
\prod_{i = 1}^{N_{Exp}} f(E;m_{\chi}, \sigma_{\chi-p})
\label{likelihood}
\end{equation}
where
\begin{equation}
f(E;m_{\chi}, \sigma_{\chi-p}) =
\frac{dN/dE(E;m_{\chi}, \sigma_{\chi-p})}
{\int_{E_{th}}^{E_{sup}} dN/dE(E;m_{\chi}, \sigma_{\chi-p})}
\end{equation}
is the normalized \textit{total} event rate (signal+background) and $N_{th}^{scan}$ is the theoretical number of events, expected from Eq.(\ref{Recoil}), for the
given point of the parameter space. The normalization renders $f$ a probability
density function and, thus, suitable for use in a likelihood calculation.

The use of equation (\ref{likelihood}) presents the advantage that it takes into account the fact that the number of observed
events in an experiment can, actually, deviate from the expected behaviour for several reasons. For the given
experiment, say $j$, we scan over the $(m_{\chi}, \sigma_{\chi-p})$ parameter space and calculate the value
$(m_{\chi}^{Est, j}, \sigma_{\chi-p}^{Est, j})$ that maximize the expression (\ref{likelihood}). This is the estimation for the $j$-th
experiment.
We then calculate the mean value of all the estimations and find which experiment's estimation was closest to this mean value.
This experiment is considered to be the most representative of them all and is used to perform a final scan.
Finally, from the likelihood distribution we obtain through this scan we can plot discrimination capacity regions.

Direct Detection experiments present the advantage of quite well-controlled background. The additional ambiguity that arises
in Indirect Detection and concerns uncertainties in the background will be dealt with in the relevant chapter.

As a final remark on the statistical treatment we used, let us say
that in order to be more precise, we would have to take into account (as is systematically
done in \cite{Baltz:2008wd}) the fact that the mass and cross-section precision are
themselves random variables and should, consequently, be given with their relevant
statistical variance. To do so, we would have to consider the actual distribution of
estimators for all experiments. However, as discussed in the Introduction,
such a treatment goes far beyond the scopes of this paper, where we are interested
in a more qualitative comparison of different detection modes. In this respect, we
keep the experiment which averages the properties of a larger set of
experiments. Motivating this approach, our results are indeed in accordance with \cite{Baltz:2008wd} and
\cite{Green:2008rd}.

\subsection{The XENON experiment}

\begin{figure}[!]
    \begin{center}
\centerline{
%\vskip -4.truecm
       \epsfig{file=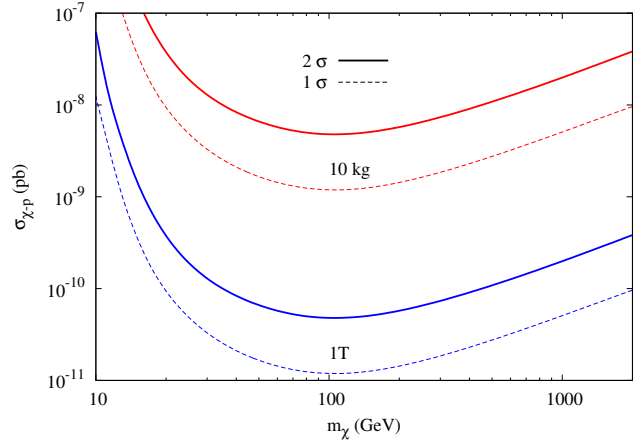,angle=-90,width=0.5\textwidth}
       }
          \caption{{\footnotesize
Spin-independent WIMP-nucleon cross-section versus WIMP mass
for $\chi^2=1,4$  and M=10 kg and 1 ton.
}}
        \label{fig:Xenonsensitivity}
    \end{center}

\end{figure}

The XENON experiment at the Gran Sasso national laboratory aims
at the direct detection of dark matter via its elastic scattering off
xenon nuclei.
It was deployed underground in March $2006$ and has been in continuous operation
for a period of about one year. It allows the simultaneous measurement
of direct scintillation in the liquid and of ionization,
via proportional scintillation in the gas. In this way, XENON discriminates
signal from background for a nuclear recoil energy as small as $4.5$ keV.
Currently a $10$ kg detector is being used, but the final
mass will be $1$ ton of liquid xenon.
In Fig.\ref{fig:Xenonsensitivity}, we show the sensitivity curve for \textit{Xenon10}
 ($M=10$ kg) and \textit{Xenon1T} ($M=1$ ton) for $T=3$ years of data acquisition,
supposing zero backgrounds and a perfectly known velocity of the sun
around the galactic center, fixed at $220$ km/sec.

In our study, following Ref. \cite{Angle:2007uj}
we will consider the energy range between $4$ and $30$ keV and $3$ years of data acquisition for
a $100$ kg XENON experiment.  Such experimental conditions and time
of exposure can be achieved after the $6$ years of GLAST mission and justify
the comparison between the two detection modes.

\begin{figure}[!]
    \begin{center}
\centerline{
%\vskip -4.truecm
       \epsfig{file=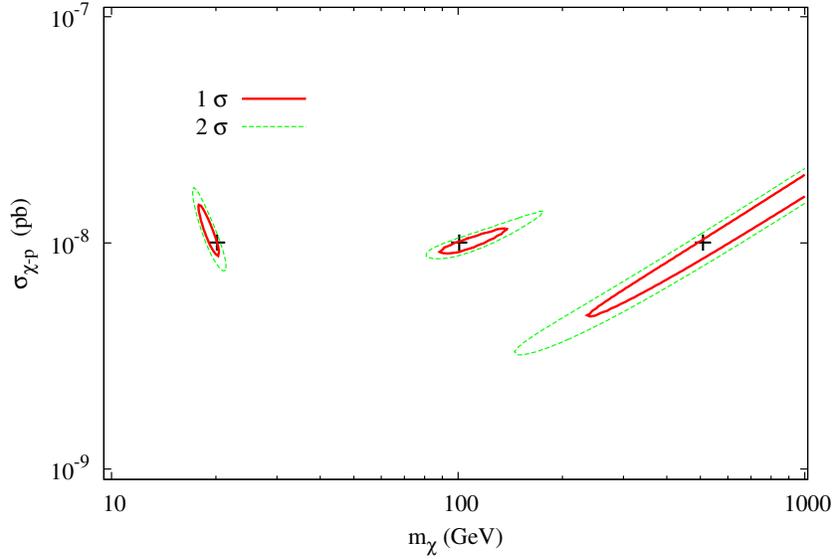,width=0.45\textwidth,angle=-90}
       }
          \caption{{\footnotesize
Distribution of the maximum likelihood WIMP mass, $m_{\chi}$, and
cross-section, $\sigma_{\chi-p}$, for $3$ years of exposure in a $100$ kg
XENON experiment, for
$m_{\chi}=20,~100$, $500$ GeV and $\sigma_{\chi-p}=10^{-8}$ pb.
The inner (full) and outer (dashed) lines represent the $68\%$ and
$95\%$ CL region respectively.
The crosses denote the theoretical input parameters
($\sigma_{\chi-p}$, $m_{\chi}$).
}}
        \label{fig:Direct}
    \end{center}

\end{figure}

In Fig.\ref{fig:Direct} we show
the ability of XENON to determine the mass and scattering
cross-section for a $20$ , $100$ and $500$ GeV
WIMP. We can clearly see how sensitive the experiment is to
light WIMPs: the precision can easily reach the percent level for
\footnote{During
the finalization of this work, Drees and Shan in Ref.  \cite{Green:2007rb}
proposed that one can even increase such a precision with a combined analysis of
two experiments of direct detection.} $m_{\chi}\lesssim 50$ GeV. Indeed, the recoil energy of the
nucleus depends on the reduced mass (see Eq.(\ref{Recoil})). For
WIMPs much heavier than the nucleus mass ($\sim 100$ GeV for Xenon),
$m_r \sim m_N$, and is therefore independent of the WIMP mass.
This is clearly reflected in the uncertainties
at $68\%$ and $95\%$ CL in Fig.\ref{fig:Direct} for a 500 GeV WIMP.

\subsection{Influence of astrophysical/background assumptions}
\label{Directuncertainties}
As mentioned before, a significant uncertainty (of the order of $8-10$\%) exists for the largely
used values of the sun's circular velocity around the Galactic Center (GC), as well
as the various background
forms we could expect in direct detection experiments. In this respect, it would be
interesting to examine how the previous results are altered in the case where $v_0$ is actually
included in the fitting procedure, letting it vary within the given margin of error.

As far as background events are concerned, it is quite difficult to perform a general
study valid for every detector. Neutron backgrounds, which are in fact the most
difficult to distinguish from signal events, are usually taken to come from three sources
(see also \cite{Aprile:2002ef}):
\begin{itemize}
\item Cosmic muon - induced neutrons, which are not in general considered to cause
much nuisance.
\item Neutrons from the detector's surrounding rock.
\item Neutrons coming from contamination of the detector itself or surrounding
\end{itemize}
materials.

As we said, it is difficult to model in general neutron backgrounds, as they are
mostly determined by the specific location in which every experiment is situated, as
well as by the specific shielding configuration adopted by each collaboration.
Two widely studied forms of neutron backgrounds are the case of a constant one,
which seems to be quite well-motivated by an experimental point of view and can
resemble to a heavy WIMP's signal, and an
exponential one which apart from its theoretical motivation is also interesting
as it gets to ``mimic'' (as pointed out in \cite{Green:2008rd}) the actual signal spectrum
for intermediate WIMP masses.
In this respect, we studied the impact of these two forms of background:

We consider firstly a constant background, with a value taken to be the
same as the maximal WIMP signal
in the first energy bin.
Throughout this paper, when examining the impact of uncertainties on
the mass determination accuracy, we will consider the case of a
somehow ``typical'' in many theoretical frameworks
case of a 100 GeV WIMP.

Then, we introduce an exponential background of the form
$\left (\frac{dN}{dE}\right )_{\mbox{bkg}}= A  \exp(-E/E_b)$,
where the slope of the exponential is fixed at $E_b = 25$ keV and
the $A$ factor is determined by demanding that the maximal
values of the signal and the background be the same. The
reason for this specific choice of parameters is that it is
for these values that the signal spectrum has a significant
resemblance to the background one, making it difficult to distinguish
from one another.

\begin{figure}[!]
    \begin{center}
\centerline{
%\vskip -4.truecm
       \epsfig{file= 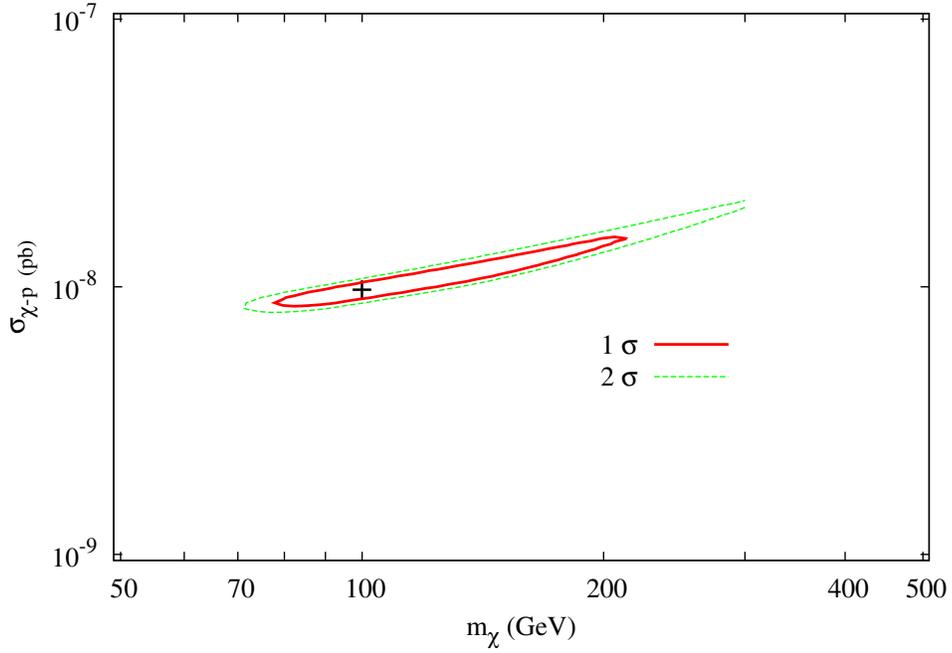,width=0.45\textwidth,angle=-90}
       }
\vspace{1 cm}
          \caption{{\footnotesize
$68\%$ and $95\%$ CL regions for the XENON $100$ kg experiment for a $100$ GeV WIMP
with a proton-WIMP scattering cross-section of $10^{-8}$pb
in the case where uncertainties in the $v_0$ parameter are taken into
account and, thus, included in the fitting procedure.
}}
        \label{fig:XENONv0}
    \end{center}

\end{figure}

\begin{figure}[!]
    \begin{center}
\centerline{
%\vskip -4.truecm
       \epsfig{file=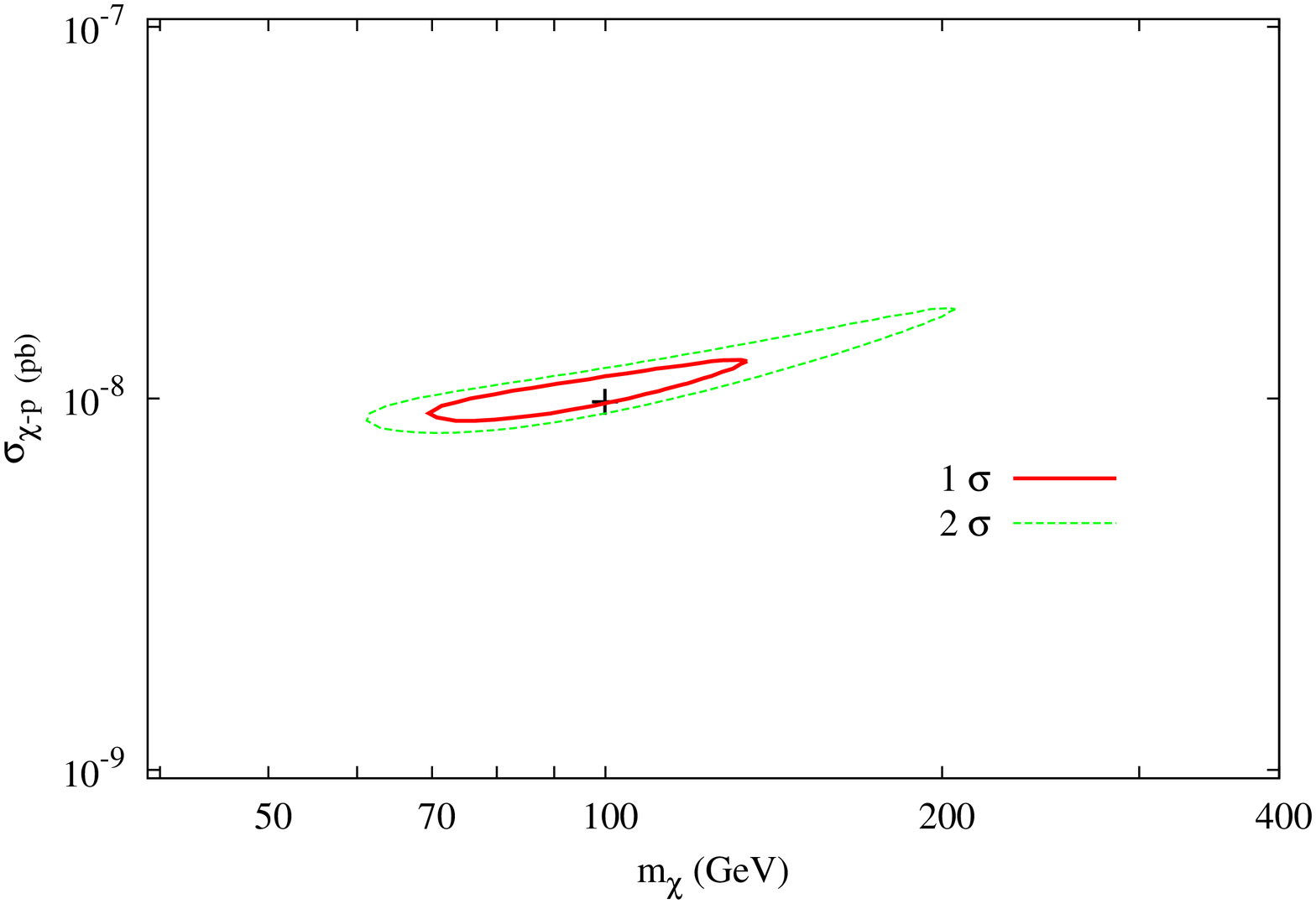,width=0.3\textwidth,angle=-90}
\hspace{0.5 cm}
      \epsfig{file=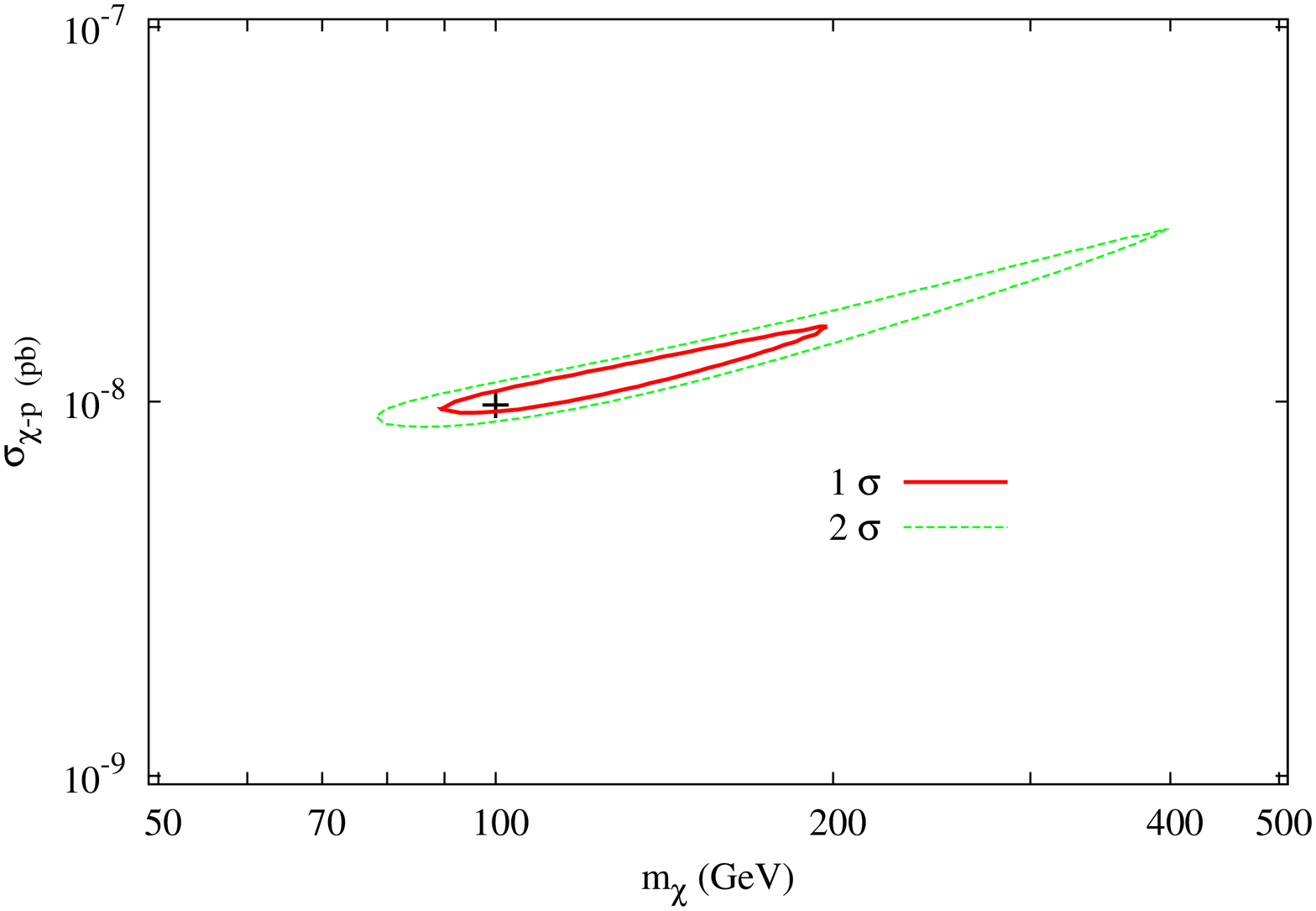,width=0.3\textwidth,angle=-90}
       }
\vspace{1 cm} %\\
          \caption{{\footnotesize
$68\%$ and $95\%$ CL regions for the XENON $100$ kg experiment for a $100$ GeV WIMP
with a proton-WIMP scattering cross-section of $10^{-8}$pb
including a constant neutron background (left)
or an exponential one (right). The serious deterioration of accuracy in the
second case is due to the fact that the background parameters where chosen
in order to mimic the actual signal spectrum.
}}
        \label{fig:XENONbkg}
    \end{center}

\end{figure}

Our results are shown in Figs.\ref{fig:XENONv0} and \ref{fig:XENONbkg}
for the cases of inclusion of $v_0$ in the fitting procedure and non-zero
backgrounds respectively.
The deterioration of the expected accuracy is obvious, when we compare these plots
to those of Fig.\ref{fig:Direct}. Especially for the case of large uncertainties
in $v_0$ (we let it vary in the whole margin-of-error region) and of inclusion
of a background which is nearly degenerate with the signal, the expected precision
is dramatically aggravated. This shows, among others, the extreme importance of
a well-controlled environment and well-measured input parameters, other than the
WIMP mass.

%%%%%%%%%%%%%%%%%%%%%%%%%%%%%%%%%%%%%%%%%%%%%%%%%%%%%%%%%%%%%%%%%%%%%%%%%%%%%%%%%
%%%%%%%%%%%%%%%%%%%%%%%%%%%	INDIRECT DETECTION %%%%%%%%%%%%%%%%%%%%%%%%%%%%%%%%%%
%%%%%%%%%%%%%%%%%%%%%%%%%%%%%%%%%%%%%%%%%%%%%%%%%%%%%%%%%%%%%%%%%%%%%%%%%%%%%%%%%

\section{Indirect detection}

\subsection{Differential event rate}

The spectrum of gamma--rays generated in dark matter annihilations
and coming from a direction forming an angle $\psi$ with respect to
the galactic center is
\begin{equation}
%\frac{d \Phi_{\gamma}}{d \Omega d E}
\Phi_{\gamma}(E_{\gamma}, \psi)
=\sum_i
%\frac{1}{2}
\frac{dN_{\gamma}^i}{dE_{\gamma}}
 Br_i \langle\sigma v\rangle \frac{1}{8 \pi m_{\chi}^2}\int_{line\
 of\ sight} \rho^2
% (r(l,\psi))
\ dl\ ,
\label{Eq:flux}
\end{equation}
where the discrete sum is over all dark matter annihilation
channels,
$dN_{\gamma}^i/dE_{\gamma}$ is the differential gamma--ray yield,
$\langle\sigma v\rangle$ is the annihilation cross-section averaged
over its velocity distribution, $Br_i$ is the branching ratio of annihilation
into final state ``i'' , and $\rho$ is the dark matter density.
The method followed in order to obtain the
spectral function describing the standard model particle decay into
$\gamma$-rays, is presented in the Appendix.

\begin{center}
\begin{table}
\centering
\begin{tabular}{|c|ccccc|}
\hline
&$a$ (kpc)&$\alpha$&$\beta$&$\gamma$
%&$\bar{J}(10^{-3} {\rm sr}) $
&$\bar{J}(4\cdot10^{-3} {\rm sr})$  \\
\hline
NFW & $20$ & $1$ & $3$ & $1$
%& $1.214 \cdot 10^3$
& $5.859\cdot10^2$\\
$\rm{NFW_c}$ & $20$ & $0.8$ & $2.7$ & $1.45$
%& $1.755  \cdot 10^5$
& $3.254\cdot10^4$\\
Moore et al. & $28$ & $1.5$ & $3$ & $1.5$
%& $1.603  \cdot 10^5$
& $2.574\cdot10^4$\\
$\rm{Moore_c}$ & $28$ & $0.8$ & $2.7$ & $1.65$
%& $1.242  \cdot 10^7$
& $3.075\cdot10^5$\\
\hline
\end{tabular}
\caption{{\footnotesize NFW and Moore et al.
density profiles without
and with
adiabatic compression ($\rm{NFW}_c$ and $\rm{Moore_c}$ respectively)
with the corresponding parameters, and values of $\bar{J}(\Delta\Omega)$.}}
%for $\Delta\Omega=10^{-3}, 10^{-5}\ {\rm sr}$.}
\label{tab}
\end{table}
\end{center}

It is customary to rewrite Eq.~(\ref{Eq:flux}) introducing the
dimensionless quantity $J$ (which depends only on the dark
matter distribution):
\begin{equation}
J(\psi)=\frac{1}{8.5 ~\mathrm{kpc}}
\left(
\frac{1}{0.3 ~\mathrm{GeV/cm^3}}
\right)^2
\int_{line\ of\ sight}\rho^2(r(l,\psi))\ dl\ .
\label{Jbarr}
\end{equation}
After having averaged over a solid angle, $\Delta \Omega$,
the gamma--ray flux can now be expressed as
\begin{eqnarray}
\Phi_{\gamma}(E_{\gamma})
& = &
0.94\cdot 10^{-13}\ \mathrm{cm^{-2}\ s^{-1}\ GeV^{-1}\ sr^{-1}}
\nonumber \\
&\cdot & \mbox{}
\sum_i
\frac{dN_{\gamma}^i}{dE_{\gamma}}
\left(
\frac{Br_i\langle\sigma
v\rangle}{10^{-29} {\mathrm{cm^3 s^{-1}}}}
\right)
\left(
\frac{100 ~\mathrm{GeV}}{m_{\chi}}
\right)^2
{\overline{J}}(\Delta \Omega) \Delta \Omega \ .
\label{Eq:totflux}
\end{eqnarray}
The value of $\overline{J}(\Delta \Omega) \Delta \Omega$ depends
crucially on the dark matter distribution.
The most common parametrization of the different profiles
that have been proposed in the literature is

\begin{equation}
\rho(r)= \frac{\rho_0  [1+(R_0/a)^{\alpha}]^{(\beta-\gamma)/\alpha}  }{(r/R_0)^{\gamma}
[1+(r/a)^{\alpha}]^{(\beta-\gamma)/\alpha}}\ ,
\label{profile}
\end{equation}
where $\rho_0$ is the local (solar neighborhood)
halo density,
% ($\simeq$ 0.3 GeV/cm$^3$),
$a$ is a characteristic length, and $R_0$ the distance from the Sun
to the galactic center.
As mentioned above, we will use $\rho_0=0.3$ GeV/cm$^3$ throughout the paper, but
since this is just a scaling factor in the analysis,
modifications to its value can be straightforwardly taken into account
in the results.
N--body simulations suggest a cuspy inner region of dark
matter halo with a distribution
where $\gamma$ generally lies in the range $1$ (NFW profile \cite{Navarro:1995iw})
to $1.5$ (Moore et al. profile \cite{Moore:1999nt}),
producing a profile with a behavior $\rho(r) \propto r^{-\gamma}$ at small distances.
Over a solid angle of
$4\cdot10^{-3}$ sr, such profiles can lead from
$\overline{J}(\Delta \Omega)\sim 5.859\cdot10^2$
to $2.574\cdot10^4$. Moreover, if we take into account the baryon distribution
in the Galaxy, we can predict even more cuspy profiles with
$\gamma$ in the range $1.45$ to $1.65$
($\overline{J}(\Delta \Omega) \sim 3.254\cdot10^4 - 3.075\cdot10^5$) through
the adiabatic compression process
(see the study of Refs.~\cite{Prada:2004pi,Mambrini:2005vk}).
We summarize the parameters used in our study and the values of
$\overline{J}$ for each profile in Table \ref{tab}.
The values contained in the Table were calculated using the darkSUSY
package for calculations of fluxes coming from SUSY DM candidates' annihilations.
The calculation is obviously not altered whatever DM candidate is
assumed and is therefore valid for an arbitrary WIMP as the ones examined here.

It is worth noticing here that we are neglecting the effect of
clumpyness,
even though other studies showed that, depending
upon assumptions on the clumps' distribution, in principle an enhancement
of the flux by a factor of $2$ to $10$ is possible \cite{Clumps2}. In this respect, the
following predictions on the gamma-ray flux from the galactic center are
conservative.

\subsection{Modeling the galactic center background.}

HESS \cite{Aharonian:2004wa} has measured the
gamma--ray spectrum from the galactic
center in the range of energy $\sim$ [$160$ GeV--$10$ TeV]. The collaboration
 claims that the data are fitted by a power--law
\begin{equation}
\phi^{\mathrm{HESS}}_{\mathrm{bkg}}(E) = F_0 ~ E_{\mathrm{TeV}}^{-\alpha},
\end{equation}

\noindent
with a spectral index
$\alpha=2.21 \pm 0.09$ and
$F_0=(2.50 \pm 0.21) \cdot 10^{-8} ~\mathrm{m^{-2} ~ s^{-1} ~ TeV^{-1}}$.
The data were taken during the second phase of measurements
(July--August, $2003$) with a $\chi^2$ of $0.6$ per degree of
freedom. Because of the constant slope
power--law observed by HESS, it turns out possible but difficult to
conciliate such a spectrum with a signal from dark matter annihilation
\cite{Mambrini:2005vk, Profumo:2005xd}.
Indeed, final particles (quarks, leptons or gauge bosons)
produced through annihilations give rise to a spectrum with a continuously
changing slope. Several astrophysical models have been proposed in order
to match the HESS data \cite{Aharonian:2004jr}.
In the present study we consider the astrophysical background for
gamma--ray detection as the one extrapolated from the HESS data with
a continuous power--law over the energy range of interest
($\approx 1$ -- $300$ GeV).
As was recently underlined in Ref. \cite{Zaharijas:2006qb},
the sensitivity of GLAST will be affected by the presence of such an
 astrophysical source.
Note that the WIMP masses that we shall obtain in our parameter space
$\lesssim 1$ TeV avoid any conflict with the observations of HESS.

In addition,
we have also taken into account the EGRET data \cite{EGRET} in our background
at energies below $10$ GeV ($\phi^{\mathrm{EGRET}}_{\mathrm{bkg}}(E)$),
as they can affect the sensitivity of the analysis.
Indeed, the extrapolation of the gamma--ray fluxes measured
by HESS down to energies as low as $1$ GeV is likely to be an underestimation
of the gamma--ray background in the galactic center, as EGRET measurements
are one to two orders of magnitude higher than the HESS extrapolation.
The EGRET point-source has been found \cite{Dodelson:2007gd} to be well
fitted by a simple power-law with slope $-2.2$.
We thus decided to take as background an interpolation between the HESS
extrapolation and the EGRET data below $10$ GeV to stay as conservative
as possible in evaluating the gamma--ray background.

Finally, we will consider the diffuse background of gamma rays in the region
surrounding the galactic center.

We will describe the spectrum of the
background using the HESS observation from the Galactic Center Ridge \cite{Aharonian:2004wa},
which can be described by

\begin{equation}
\phi^{\mathrm{diff}}_{\mathrm{bkg}}(E) = 1.1 \cdot 10^{-4}
E_{\mathrm{GeV}}^{-2.29} \mathrm{GeV^{-1} cm^{-2} s^{-1} sr^{-1}}\ .
\end{equation}

\noindent
In our analysis, we will consider the inner $2^o \times 2^o$ field of
view ($\Delta\Omega=4\cdot 10^{-3}$ sr) and the energy region between
$1$ and $300$ GeV.
During the completion of our work,
the authors of Ref. \cite{Dodelson:2007gd} gave a more detailed
and sophisticated statistical analysis
of the diffuse background, adopting an overall normalization around
the galactic center and taking into account a statistical spread
function as GLAST would be able to probe in this region.
However, we have checked
that our results are not significantly modified and we recover similar results
concerning the prospects of GLAST.

\subsection{The GLAST experiment}

\begin{figure}[!]
    \begin{center}
\centerline{
%\vskip 4.truecm
       \epsfig{file=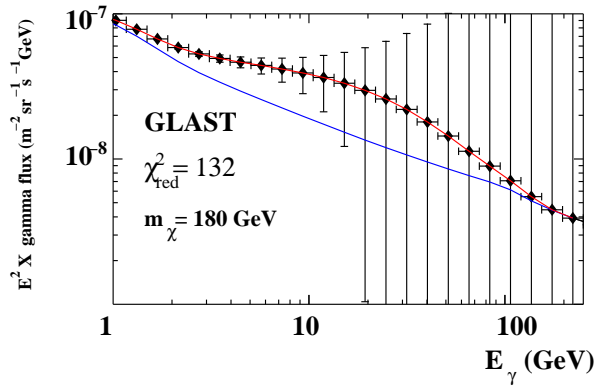,width=0.45\textwidth}
       }
          \caption{{\footnotesize
GLAST expectations for gamma--ray fluxes in the case of
 a WIMP mass $m_{\chi}=180$ GeV and cross section
$\langle\sigma v\rangle = 3\cdot 10^{-26}$ cm$^3$s$^{-1}$.
A NFW halo profile has been adopted.
The error bars shown are those projected for the GLAST experiment
after a six-year mission run,
assuming the galactic center will be within its field--of--view
$50\%$ of the time. The lower (blue) line is the
background--only prediction.
The $\chi^2_{red}$ is 132, giving
a signal clearly distinguishable from the background.
}}
        \label{fig:GLAST}
    \end{center}

\end{figure}

The space--based gamma--ray telescope GLAST \cite{GLAST} was
launched in May $2008$ for a five-year mission. It
will perform an all-sky survey covering a large energy range
($\approx 1$ -- $300$ GeV). With an effective area and angular
resolution of the order of $10^4 ~ \mathrm{cm^2}$ and
$0.1^o$ ($\Delta \Omega \sim 10^{-5}$ sr) respectively,
GLAST will be able to point and analyze the inner center of the Milky
Way ($\sim 7$ pc).
Concerning the
%requested condition on the $\chi^2$
statistical method,
%for a signal discovery,
we have used an analysis similar to the one considered in the
case of direct detection in
section \ref{statistical}, with a six-year mission run,
assuming the galactic center will be within its field--of--view
$50\%$ of the time \cite{GLAST,Gehrels:1999ri}.
In Fig.\ref{fig:GLAST} we show the ability of GLAST to identify
a signal from dark matter annihilation for a WIMP mass
of $180$ GeV.
The error bars shown are projected assuming Gaussian statistic, and
we adopt the background described above including Poisson noise.
In the following, we will concentrate on a process
which gives $100\%$ annihilation to $WW$. We have checked that the dependence
on the final state does not influence significantly the general
results of the study, except in the case of leptonic final states. This
will be studied in a specific case in Section 5 dedicated to the  ILC
experiment.

\begin{figure}[!]
    \begin{center}
\centerline{
%\vskip -4.truecm
%       \epsfig{file=DirectSigmam8.eps,width=0.45\textwidth,angle=-90}
       \epsfig{file=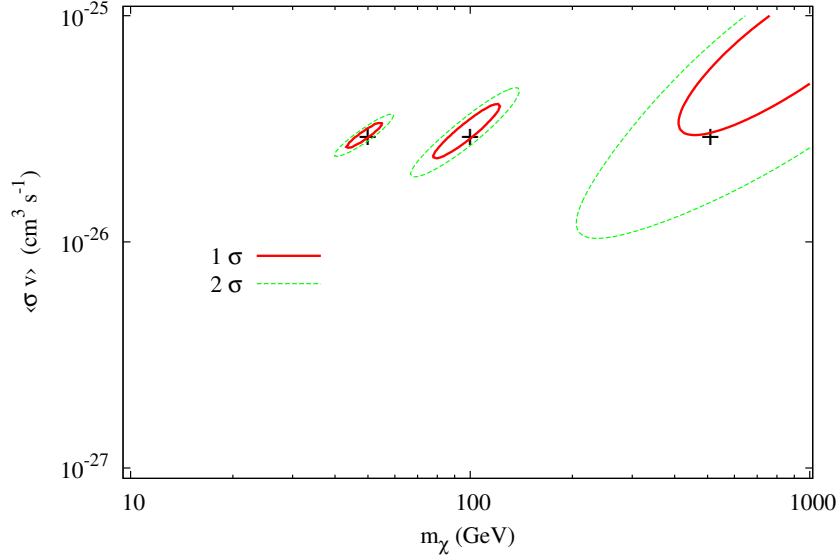,width=0.45\textwidth,angle=-90}
       }
          \caption{{\footnotesize
Distribution of the maximum likelihood WIMP mass, $m_{\chi}$, and
annihilation cross-section, $\langle\sigma v\rangle$, after 6 years of observation
($50\%$ of time exposure)
of the galactic center with GLAST, with the hypothesis of a NFW halo profile,
for $m_{\chi}=50,~100$, $500$ GeV and
$\langle\sigma v\rangle=3 \cdot 10^{-26}$ cm$^3$s$^{-1}$.
The inner (full) and outer (dashed) lines represent the $68\%$ and
$95\%$ CL region respectively.
The crosses denote the theoretical input parameters
($\langle \sigma v \rangle$, $m_{\chi}$).
}}
        \label{fig:Indirect}
    \end{center}

\end{figure}

\begin{figure}[!]
    \begin{center}
\centerline{
%\vskip -4.truecm
\includegraphics[width=0.45\textwidth,clip=true,angle=-90]{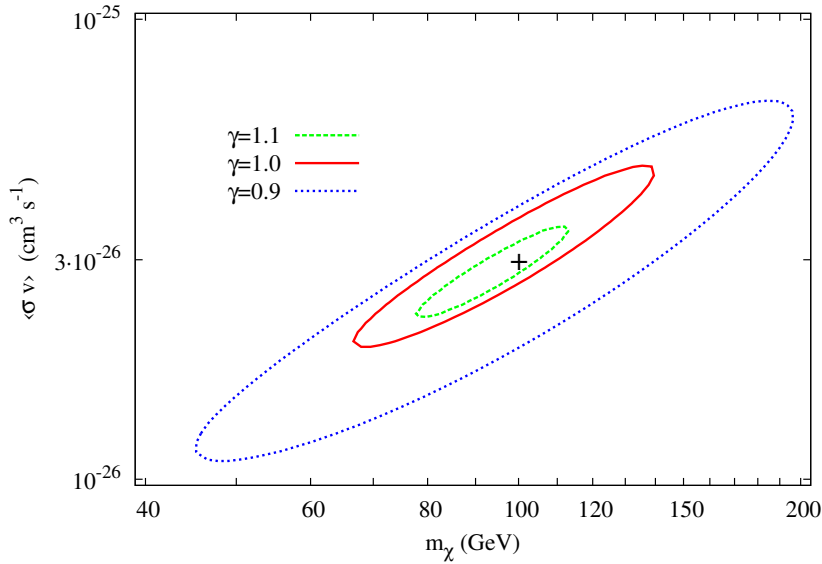}
       }
          \caption{{\footnotesize
NFW--like halo profile with $\gamma=0.9$, $1$ (NFW) and $1.1$ at $95\%$ confidence level.}}
\label{fig:profiles}
    \end{center}
\end{figure}

In Fig.\ref{fig:Indirect} we show
the ability of GLAST to determine the mass and annihilation
cross-section for a $50$, $100$ and $500$ GeV
WIMP. Again, we can see that the experiment is more sensitive to
light WIMPs: the precision can easily reach the percent level
for GLAST for $m_{\chi}\lesssim 50$ GeV.
The gamma--ray spectrum will give more precise
measurements if the mass of the WIMP lies within the GLAST
sensitivity range. Indeed, the shape of the spectrum will be easily
reconstructed above the HESS/EGRET and diffuse background if
the endpoint of the annihilation spectrum
lies within the energy range reachable by GLAST [$0.1-300$ GeV].
Furthermore, we have studied the influence of the variation of
the inner slope of the halo profile on the resolution of the WIMP mass.
In addition to the NFW profile, we have considered some NFW--like profiles,
allowing the $\gamma$ parameter in Eq.(\ref{profile})
to vary from its original value by $10\%$. This is shown in
%Thus, in
Fig.\ref{fig:profiles}, where in addition to the NFW halo profile ($\gamma=1$)
%we show the precision level given by GLAST if we assume a smoother
we also study profiles with
$\gamma=0.9, 1.1$.
As expected, the larger the $\gamma$ is, the more enhanced the galactic gamma ray flux becomes, and the better the
WIMP mass resolution turns out to be.
It is worth noticing here that in the case of a compressed NFW profile
($\gamma \sim 1.45$), the precision of GLAST increases by two orders
of magnitude.

\subsection{Varying background parameters}

\begin{figure}[!]
    \begin{center}
       \epsfig{file=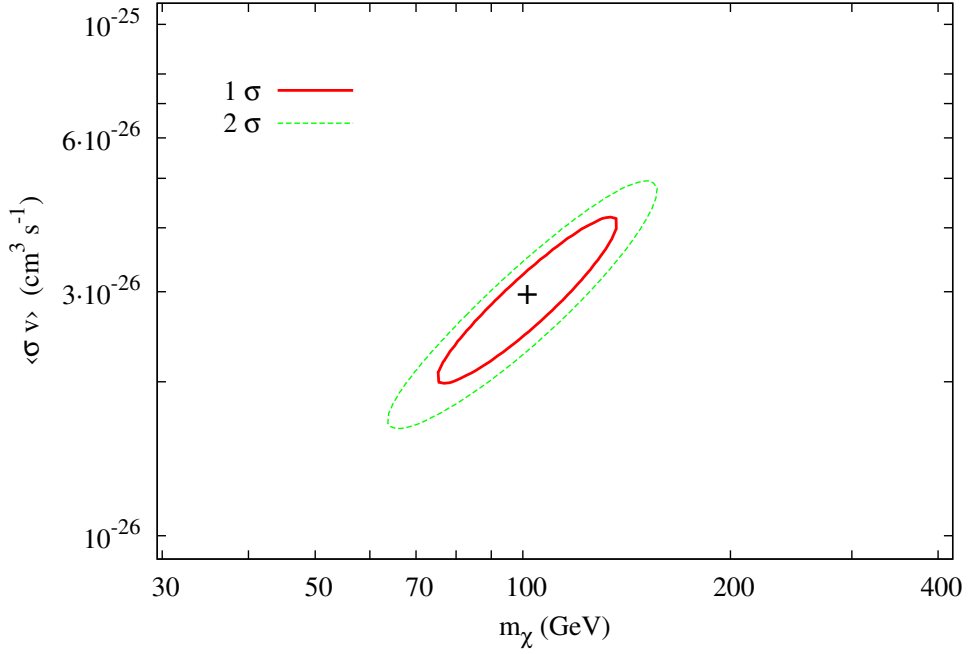,width=0.45\textwidth,angle=-90}
\vspace{1 cm} \\
          \caption{{\footnotesize
$68\%$ and $95\%$ CL regions for a statistical treatment with the
overall background normalization included in the fitting procedure, $m_\chi = 100$GeV and
$\langle\sigma v\rangle = 3\cdot10^{-26}$ cm$^3$ s$^{-1}$  .
}}
        \label{fig:TREATMENT}
    \end{center}
\end{figure}

Throughout the previous (and the following) analysis,
we have considered a perfectly known background for both detection modes.
Whereas this is a rather reasonable approximation for the case of direct detection,
it is less obvious
for the indirect one.  As it has been pointed out (see, for example,
\cite{Cesarini:2003nr, Jeltema:2008hf}), the uncertainties entering the calculation
of the backgrounds coming from the
galactic center region can considerably affect the results of any analysis.
More concretely, and especially for small WIMP masses and low energies
(where the performance of both direct
and indirect detection is maximal), the main contribution in background comes from the EGRET
source aforementioned. However, both the overall normalisation
and the spectral index characterising
this source's spectrum contain
uncertainties, and
%However,
%At the same time, at least for the case of an NFW halo profile, the
%background exceeds the signal flux by about two orders of magnitude, whereas
it has been
widely discussed in the literature that a dark matter annihilation signal could account
for some part of the EGRET source signal. An interesting point would be to include
the overall background normalization
as well as the spectral index
in the fitting procedure.
In Fig.\ref{fig:TREATMENT} we show,
for the sake of comparison,  the result of a fitting procedure, where we also fit the background
normalization while simultaneously considering signals and backgrounds
with poissonian fluctuations. The original spectrum
is taken to be the full EGRET source plus the flux produced by a $100$ GeV WIMP annihilating in a
NFW halo. One could imagine discarding low-energy data which contain a maximal background
contamination. This, however, would significantly reduce the statistics and the corresponding
precision, since a major part of the signal would be discarded.
The fact that the inclusion of an uncertainty in the background normalization (i.e. its
inclusion in the statistical treatment) does not have a major impact on the results
can be explained from the fact that throughout this work we have used  eq.(\ref{likelihood})
in our statistical analysis, which already introduces a deviation from
the ideally expected situation. In this respect, our results are quite conservative.

\begin{figure}[!]
\hspace{-3 cm}
    \begin{center}
       \epsfig{file=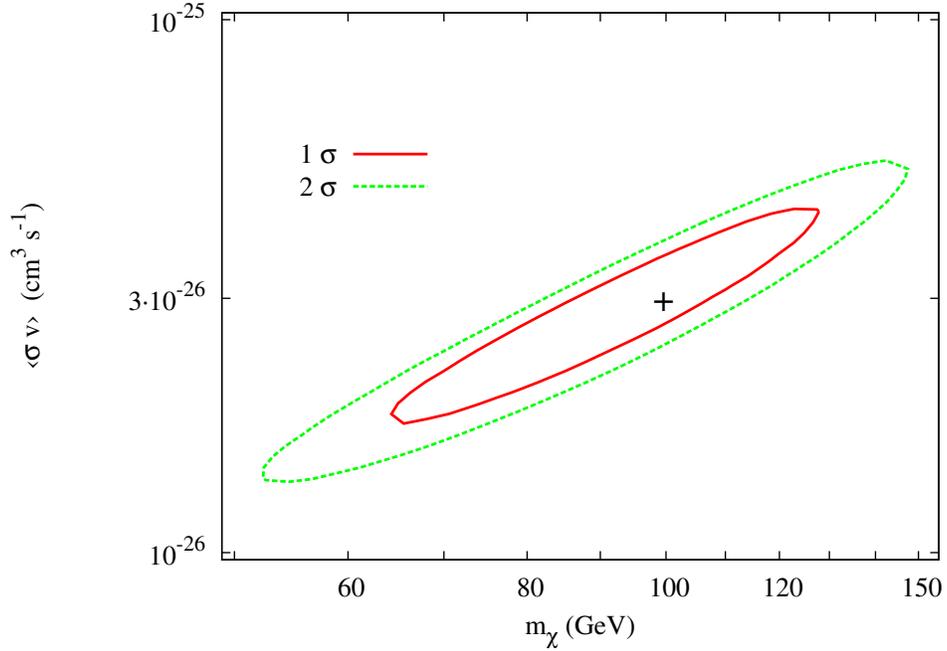,width=0.45\textwidth,angle=-90}
\vspace{1 cm} \\
          \caption{{\footnotesize
$68\%$ and $95\%$ CL regions for the case where the uncertainties in the
EGRET point source spectral index are included in the fitting procedure,
$m_\chi = 100$GeV and
$\langle\sigma v\rangle = 3\cdot10^{-26}$ cm$^3$ s$^{-1}$  . An NFW halo profile
has been assumed.
}}
        \label{fig:specindex}
    \end{center}
\end{figure}

In the same way, in Fig.\ref{fig:specindex} we show the corresponding results
where this time the spectral index is included in the fitting procedure instead.
The spectral index is left to vary in the region $[2.1, 2.4]$, which we find
to be a quite reasonable one as we verified that all over this region we obtain
reasonable fits of the EGRET data.

It is interesting to note that the variation of the background's spectral
index seems to have a larger impact on the precision that could be achieved,
with respect to the corresponding case of the background's overall
normalisation.

This is somehow logical, first of all since by definition
the background depends linearly on the normalisation factor,
but exponentially on the spectral index of the EGRET point source.
So, modifications in the latter bring along a much more drastic
modification of the background signal itsself. Furthermore, variations of
the overall normalisation have just the influence of "burying" the
signal a little more or a little less in a background which is already quite
elevated. On the contrary, by varying the spectral index we actually
change the shape of the spectrum. This brings along a \textbf{more important}
uncertainty, since we could imagine much more numerous configurations in the
(spectral index, cross-section, mass) space that could satisfy selection criteria.

\subsection{Impact of Final States}
\label{FinalStates}

Again, throughout this paper, we have considered a pure WIMP annihilation
into a $W^+ W^-$ final state. This is an assumption which is made to
simplify the overall treatment, but which at the same time somehow
restricts the generality of our results. In this paragraph, we are interested
in examining what could be the impact of variations in the final state
on the WIMP mass determination capacity. Annihlation into $ZZ$ pairs
is not expected to seriously modify the results, what would be more interesting
would be to see what happens when we consider (light or heavy) quark pairs and/or
leptons as WIMP annihilation products.

The spectrae of SM particles' annihilations into gamma-rays can be found in the
Appendix. As explained there, the only leptonic final state we consider is the
$\tau^+ \tau^-$ one, since annihilation into $\mu^+ \mu^-$ has a relatively small
contribution to the annihilation gamma-ray spectrum, whereas $e^+ e^-$ pairs
contribute through other processes, the examination of which exceeds the
purposes of the present treatment. We should, nevertheless, note here that
we do not take into account the effects of leptonic final state radiation which
can indeed become important, especially in the case of Kaluza-Klein dark matter
and in energy ranges lying near the WIMP mass.
The effect of such processes has been discussed in detail in refs.\cite{Birkedal:2005ep}
for the case of a generic WIMP and \cite{Bergstrom:2004cy} for the special case
of KK dark matter.
Obviously, this omission somehow retricts the generality of our results
as far as the impact of final states are concerned.

\begin{figure}[!]
\hspace{-3 cm}
    \begin{center}
       \epsfig{file=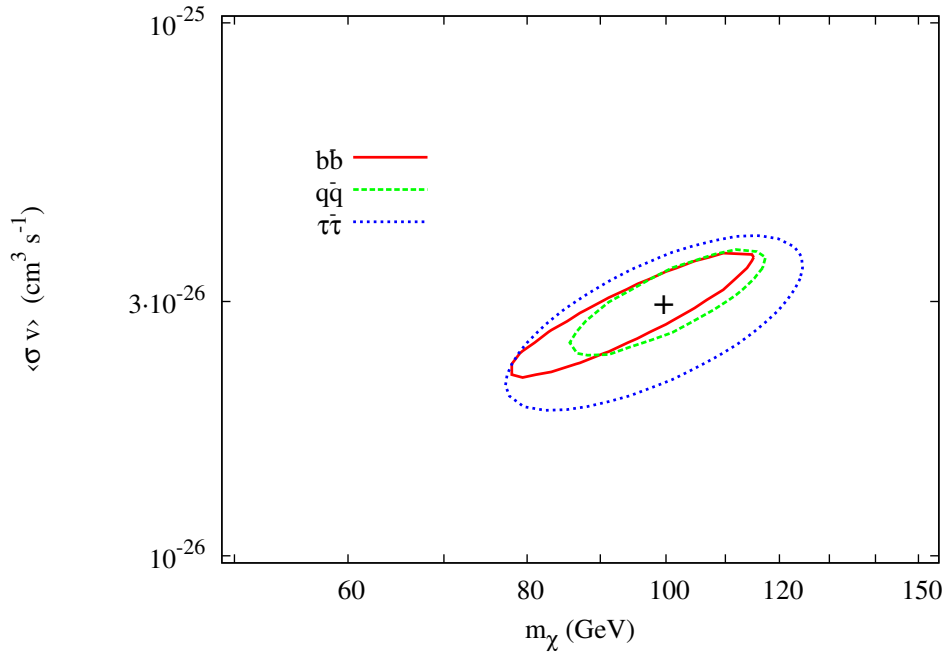,width=0.45\textwidth,angle=-90}
\vspace{1 cm} \\
          \caption{{\footnotesize
$95\%$ CL regions for a $100$ GeV WIMP and different final
states. The total thermally averaged annihilation cross-section has been taken
to be $\langle\sigma v\rangle = 3\cdot10^{-26}$ cm$^3$ s$^{-1}$.
A NFW halo profile has been assumed.
}}
        \label{fig:OtherFinalStates}
    \end{center}
\end{figure}

To this goal, we performed two kinds of tests: The first one consists only
of modifying the annihilation products, considering a perfectly known final
state (meaning that the Branching Ratios are not included in the statistical treatment).
Our results can be seen in Fig.\ref{fig:OtherFinalStates} for the case of pure
$b \bar{b}$, $q \bar{q}$ and $\tau^+ \tau^-$ final states and a $100$ GeV WIMP.

It is interesting to notice the relative amelioration of the mass resolution
with respect to the pure $W^+ W^-$ final state. A possible explanation could be that,
as can clearly be seen in gamma-ray yields presented in the Appendix, fermionic
final states tend to give more hard spectrae with respect to bosonic ones (the extreme
case being leptonic ones), rendering the spectrum more easily distinguishable from
the background. The hardest spectrum is given by the $\tau^+ \tau^-$
final state. Nevertheless, in this case the characteristic spectral form is
somewhat compensated from the reduced statistics of the signal. This is not
the case for annihilation into quarks, where the characteristic spectral form,
although less obvious than in the leptonic case,
is nevertheless combined with an important enhancement in the signal.

\begin{figure}[!]
\hspace{-3 cm}
    \begin{center}
       \epsfig{file=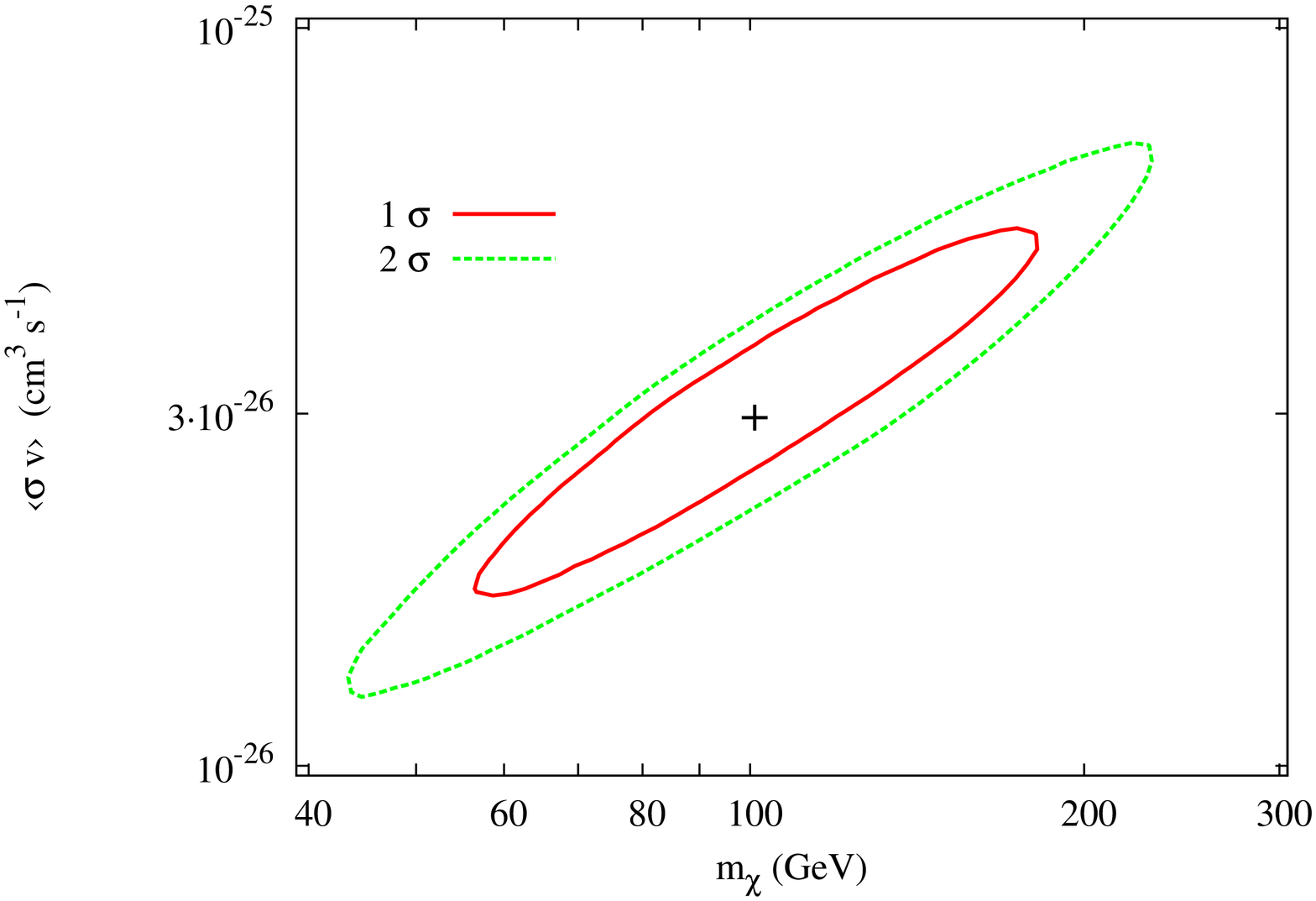,width=0.45\textwidth,angle=-90}
\vspace{1 cm} \\
          \caption{{\footnotesize
$68\%$ and $95\%$ CL regions for a $100$ GeV WIMP with
$\langle\sigma v\rangle = 3\cdot10^{-26}$ cm$^3$ s$^{-1}$
considering a mixed final state consisting of $70\%$ $W^+ W^-$ and
$30\%$ $\tau^+ \tau^-$ and including the Branching Ratios in the statistical
treatment.
}}
        \label{fig:FitBRs}
    \end{center}
\end{figure}

As a second test, we consider a mixed final state and fit the
BRs themselves along with the annihilation cross-section and the WIMP mass.
Our results can be seen in Fig.\ref{fig:FitBRs} where we have taken a 100 GeV WIMP
annihilating into a final state consisting of
$70\%$  $W^+ W^-$ and $30\%$  $\tau^+ \tau^-$. The sum of the two
branching fractions is obviously equal to $1$, so we only need to include one
further parameter in the statistical analysis.

In this case, we can clearly see that the mass resolution deteriorates
with respect to the case where a perfectly known final state is considered.
A possible explanation could be that a mixed lepton - gauge boson
final state yields a spectrum which presents neither the augmented statistics
of pure annihilation into gauge bosons (the gamma-ray yield of leptons is
significantly inferior to the one of gauge bosons) nor the characteristic hard spectral
form of annihilation into $\tau^+ \tau^-$ pairs.

%%%%%%%%%%%%%%%%%%%%%%%%%%%%%%%%%%%%%%%%%%%%%%%%%%%%%%%%%%%%%%%%%%%%
%%%%%%%%%%%%%%%%%%%%%%% COMPLEMENTARITY 1 %%%%%%%%%%%%%%%%%%%%%%%%%%
%%%%%%%%%%%%%%%%%%%%%%%%%%%%%%%%%%%%%%%%%%%%%%%%%%%%%%%%%%%%%%%%%%%%

\section{Direct versus indirect detection experiments}

It is interesting to remark from Figs.\ref{fig:Direct} and \ref{fig:Indirect},
how two completely different means of
observation, with completely different signal/background physics, are in fact
competitive (and hence complementary) in the search for the dark matter.

\begin{figure}[!]
    \begin{center}
\vskip -2.truecm
\includegraphics[width=0.45\textwidth,clip=true,angle=-90]{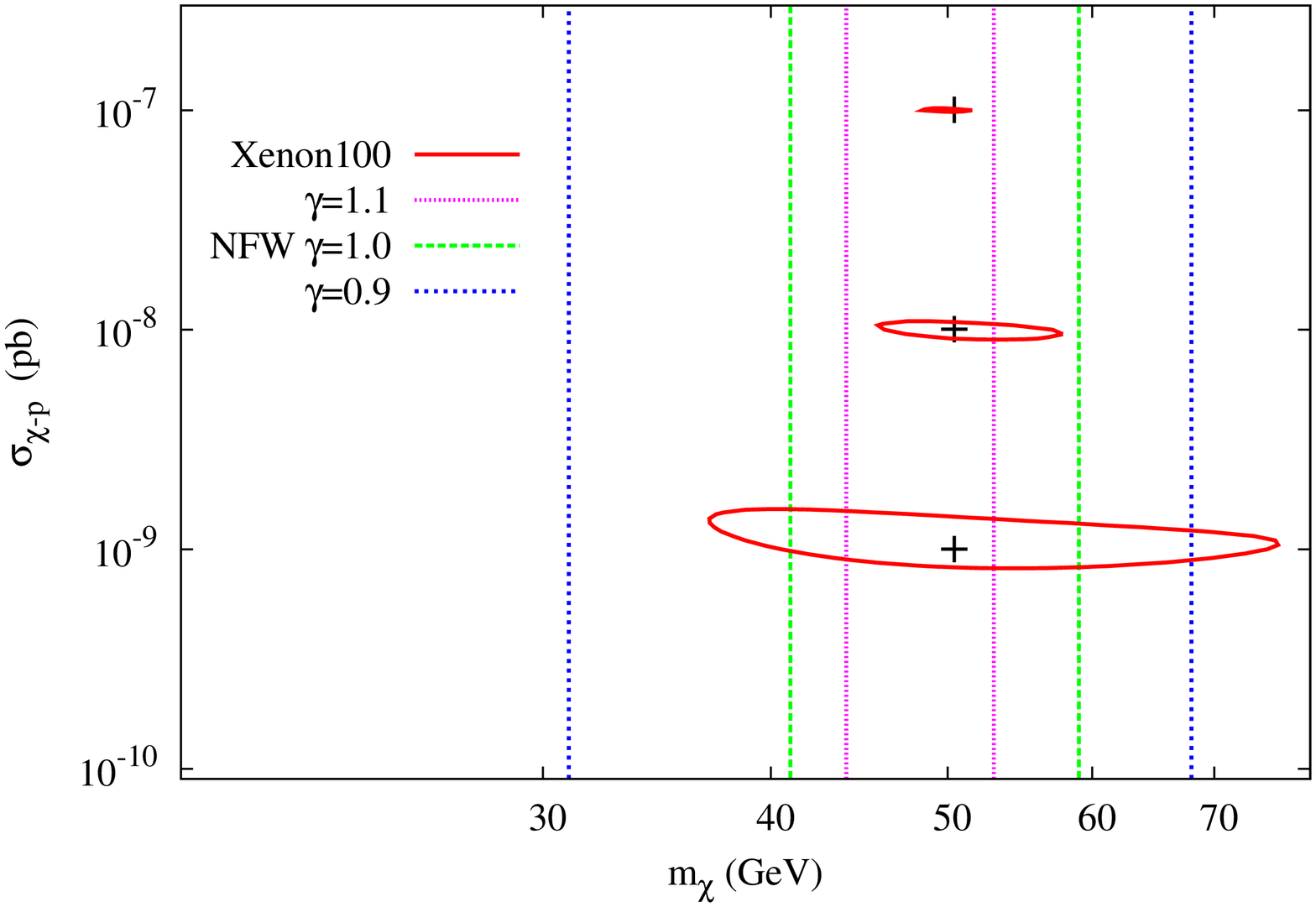}
\includegraphics[width=0.45\textwidth,clip=true,angle=-90]{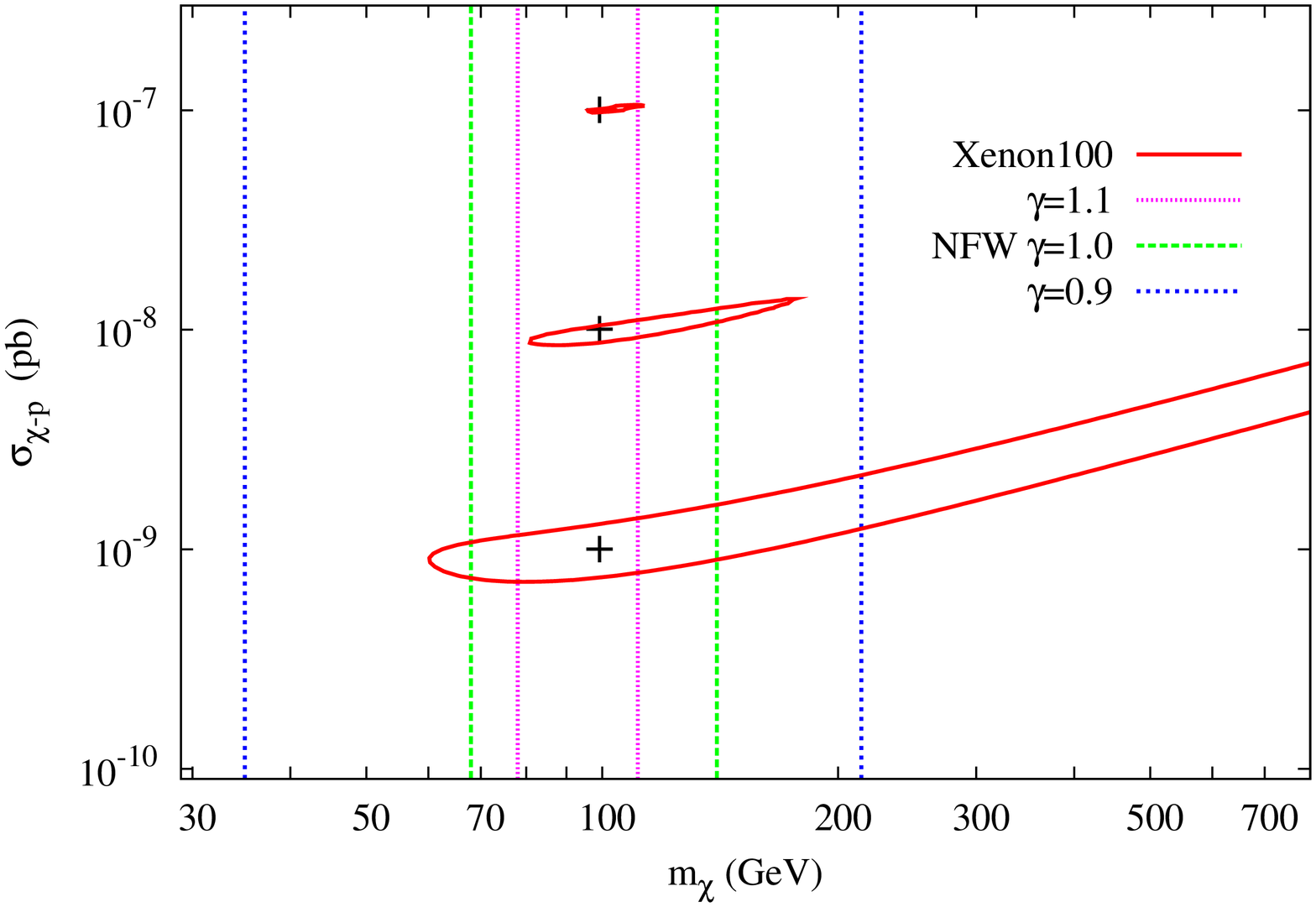}
\includegraphics[width=0.45\textwidth,clip=true,angle=-90]{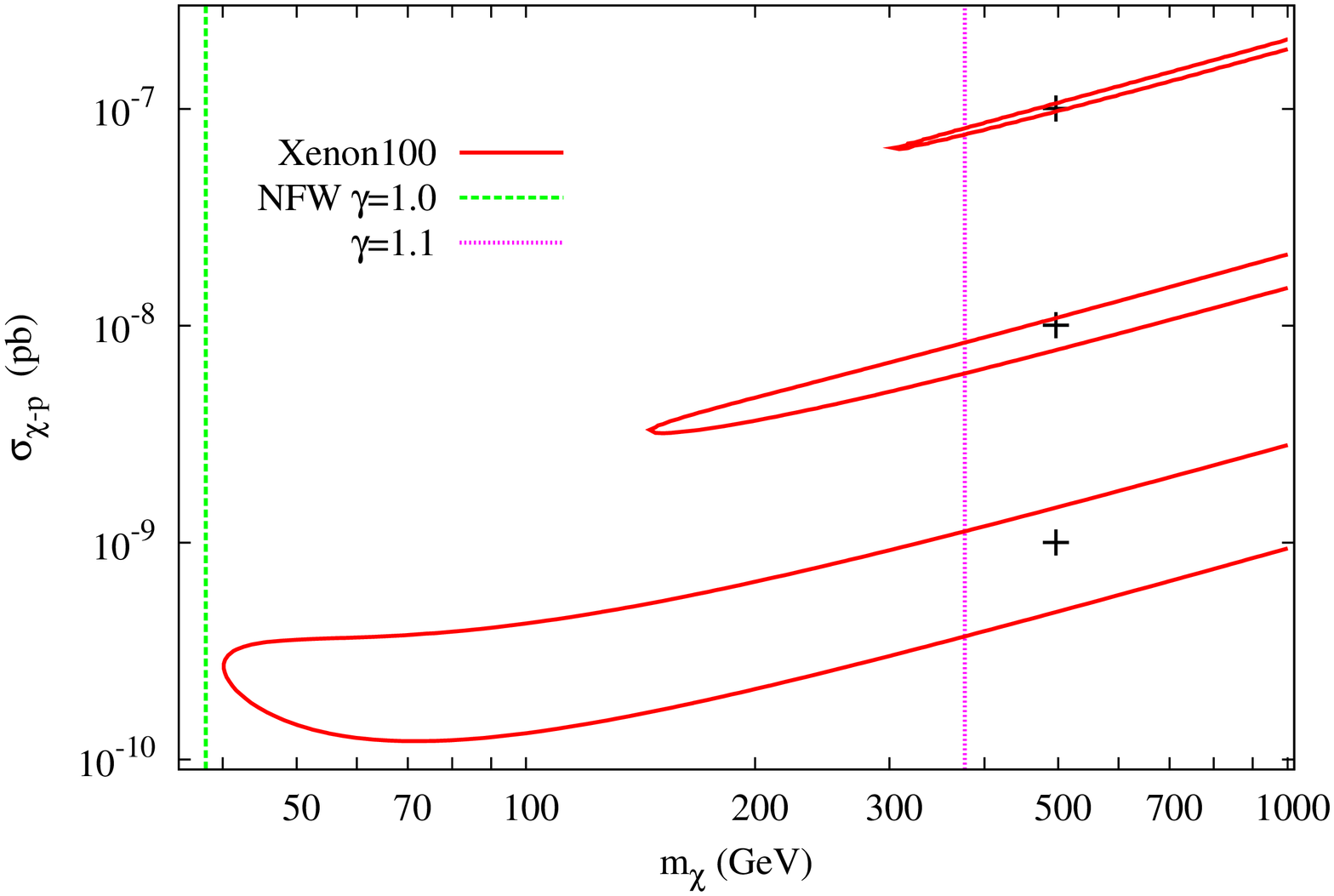}

          \caption{{\footnotesize
Comparison between $100$ kg XENON-like experiment and GLAST sensitivity
in the case of different halo profiles,
at $95\%$ of confidence level, for several WIMP masses ($50$, $100$ and $500$ GeV)
and WIMP--nucleon cross-sections ($10^{-7}$, $10^{-8}$ and
$10^{-9}$ pb).
In each panel the cross denotes the input parameters.
}}
        \label{fig:comparison}
    \end{center}

\end{figure}

In Fig.\ref{fig:comparison} we compare the precision level for both experiments
as a function of the WIMP mass, for different values of the spin-independent
cross-section ($10^{-7}$, $10^{-8}$ and $10^{-9}$ pb) and for different
halo profiles.
For this treatment we minimize the impact of uncertainties discussed in
the previous paragraph, as we are mostly interested in examining the
a priori, in some sense ``intrinsic'' sensitivity of the two detection techniques.
For example, in the case of direct detection, the necessity for minimisation
of background noises and control of the noise sources has been repeatedly stressed out.
As for uncertainties entering the velocity distribution of WIMPs in the solar
neighbourhood (or, why not, the form factor's functional form), these can,
\textit{in principle} be minimized by measurements exterior to the experiments
themselves.
The same holds for uncertainties in the case of indirect detection.
As a small example, Fermi-GLAST's overall sky survey capacity is hoped to contribute in
the minimisation of uncertainties in non-DM annihilation sources, whereas other
observations in different energy regions can also contribute in this direction.
In this respect, for our comparative results,
we remove the extra parameters from the statistical treatment.
We see
that at $95\%$ of confidence level GLAST, after $3$ years of exposure
($6$ years of taking data at $50\%$ of time exposure),
 will have an equivalent sensitivity to a $100$ kg XENON-like experiment
after $3$ years of running if $\sigma_{\chi-p}\lesssim 10^{-8}$ pb,
{\it independently of the WIMP mass}.
The indirect detection by GLAST will always be able to give an upper bound
on the WIMP mass for $m_{\chi}\sim 100$ GeV, whereas a XENON-like
$100$ kg experiment would only give a lower bound value if
$\sigma_{\chi-p} \lesssim 10^{-9}$ pb. In all cases, the lower
bounds given by GLAST for a NFW halo profile are similar to the
ones given by a XENON-like $100$ kg experiment for any WIMP mass if
$\sigma_{\chi-p} \lesssim 10^{-8}$ pb.

\begin{figure}
\begin{center}
\includegraphics[angle=-90,width=0.45\textwidth]{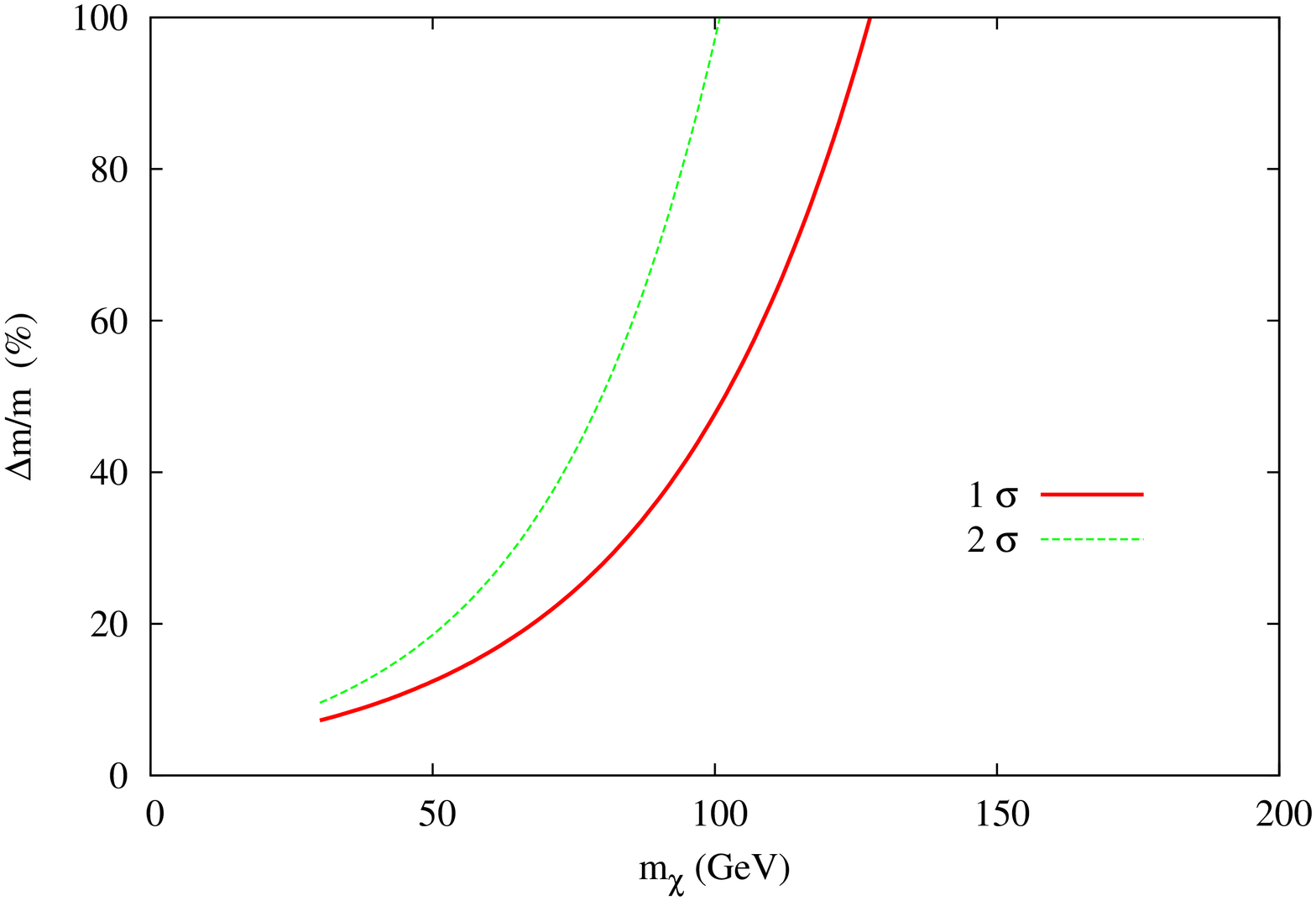}
\includegraphics[angle=-90,width=0.45\textwidth]{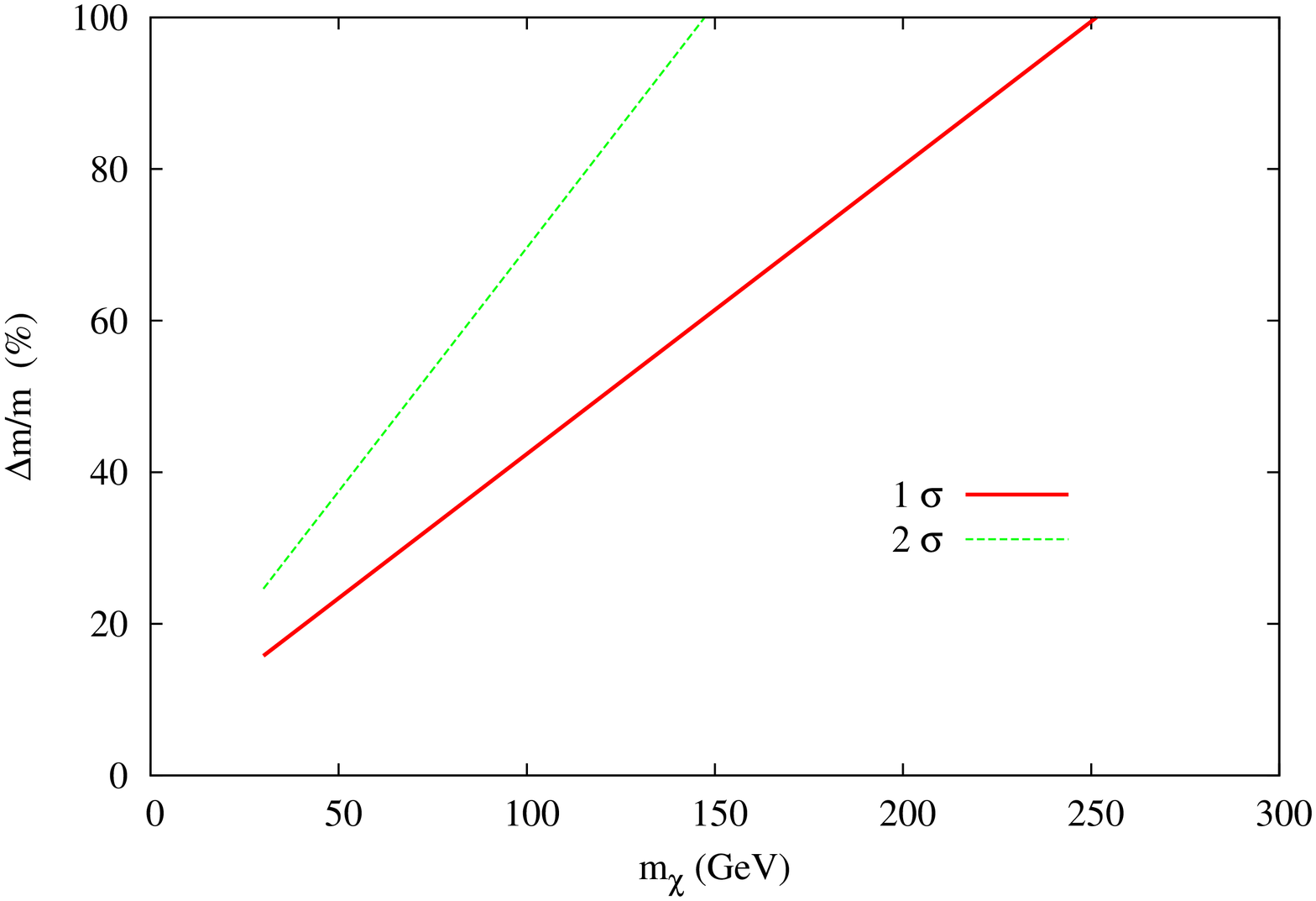}
\caption{{\footnotesize
$68\%$ and $95\%$ CL error for the XENON
$100$ kg experiment for $\sigma_{\chi-p}=10^{-8}$ pb (left) compared with the
GLAST experiment (right) for
$\langle\sigma v\rangle = 3 \cdot 10^{-26}$ cm$^3$s$^{-1}$ in the case of a NFW
halo profile.}}
\label{fig:DeltaM}
\end{center}
\end{figure}

To compare the uncertainties on the WIMP mass expected from direct
and indirect detection modes, we plotted in Fig.\ref{fig:DeltaM}
$\frac{\Delta m_\chi}{m_\chi}$
as a function of the WIMP mass for
$\sigma_{\chi-p} = 10^{-8} \mbox{pb}$ and a NFW halo profile.
One can clearly see in the figure
%left panel of Fig.\ref{fig:DeltaM}
that
GLAST will be competitive with XENON $100$ kg to measure the WIMP mass
in the case of a NFW halo profile for $\sigma_{\chi-p} \lesssim 10^{-8}$ pb.

%=========================================================================

%\begin{figure}[!]
%    \begin{center}
%%\vskip -4.truecm
%       \epsfig{file=directe.ps,angle=-90,width=0.45\textwidth}
%\hspace{0.5 cm}
%      \epsfig{file=indirecte.eps,angle=-90,width=0.41\textwidth}
%          \caption{{\footnotesize
%$68\%$ and $95\%$ CL error for the XENON
%$100$ kg experiment for $\sigma_{\chi-p}=10^{-8}$ pb (left) compared with the
%GLAST experiment (right) for
%$\langle\sigma v\rangle = 3 \cdot 10^{-26}$ cm$^3$s$^{-1}$ in the case of a NFW
%halo profile.
%}}
%        \label{fig:DeltaM}
%    \end{center}
%
%\end{figure}

%%%%%%%%%%%%%%%%%%%%%%%%%%%%%%%%%%%%%%%%%%%%%%%%%%%%%%%%%%%%%%%%%%%%
%%%%%%%%%%%%%%%%%%%%%%%%%%%%% ILC %%%%%%%%%%%%%%%%%%%%%%%%%%%%%%%%%%
%%%%%%%%%%%%%%%%%%%%%%%%%%%%%%%%%%%%%%%%%%%%%%%%%%%%%%%%%%%%%%%%%%%%

%\section{WIMP dark matter at present and future colliders}
\section{WIMPs at present and future colliders}

Among the most important sources of information concerning WIMP dark matter are,
obviously, collider experiments, both present, such as the Tevatron or the
Large Hadron Collider (LHC) and future, such as the International Linear Collider
(ILC). In fact, there is a quite
general agreement on the fact that despite the significant progress in
astroparticle physics experiments, which manage to impose more and more constraints on
various models, collider experiments remain an irreplaceable source of information
for particle physics. It is quite natural thus, to examine the potential of
colliders to constrain WIMP properties. We will examine the extent at which
astroparticle and collider experiments become competitive, trying at the same
time to stay as model-independent as possible.

This last point is, in fact, the major difficulty in treating collider experiments to extract astroparticle data:
most studies performed for new physics at colliders are very strongly model dependent. This is almost unavoidable
for the case of the LHC, due to the hadronic nature of the colliding particles. The large uncertainties on the parton
distribution functions (and, hence, on the initial energy of the colliding particles/partons) render it extremely difficult
(in fact, practically impossible) to look beyond the transverse plane. This fact obviously limits -up to a certain point- the
precision that could be obtained with respect to, for example, an $e^+e^-$ collider. As a result, it is quite difficult to make
predictions in a model-independent way, since a whole set of parameters must be taken into account in order to perform concrete
predictions. The cruciallity of these uncertainties will become clearer in the following.

\subsection{The Approach}

Recently, an approach was proposed in references \cite{Birkedal:2004xn, Birkedal:2005aa} which allows to actually
perform a model-independent study of WIMP properties at lepton colliders (such as the ILC project).
The goal we pursue is to extract constraints which are as stringent as possible for a generic dark matter candidate.
A generic WIMP can annihilate into pairs of standard model particles:
\begin{equation}
\label{annihilation}
\chi+\chi \longrightarrow X_i+\bar{X}_i\ .
\end{equation}

\noindent
However, the procedure taking place in a collider is the opposite one, having only
one species of particles in the initial state. The idea proposed in Ref. \cite{Birkedal:2004xn} is to
correlate the two processes in some way.
This can be done through the so-called ``detailed balancing'' equation, which reads:
\begin{equation}
\label{detbal}
\frac{\sigma(\chi+\chi \rightarrow X_i + \bar{X}_i)}{\sigma( X_i + \bar{X}_i \rightarrow \chi+\chi)} =
2\frac{v_X^2(2S_X+1)^2}{v_{\chi}^2(2S_{\chi}+1)^2}\ ,
\end{equation}
where $v_i$ and $S_i$ are respectively the velocity and the spin
of the particle $i$.
The cross-section $\sigma(\chi\chi \rightarrow X_i \bar{X}_i)$ is only averaged over spins.

The total thermally averaged WIMP annihilation cross-section can be expanded as
\begin{equation}
\label{expansion}
 \sigma_i v = \sum_{J=0}^{\infty}{\sigma_i^{(J)}v^{2J}}\ ,
\end{equation}
where $J$ is the angular momentum of each annihilation wave. Now, for low velocities, the lowest-order
non-vanishing term in the last equation will be dominant. So, we can express the total annihilation cross-section
as a sum of the partial ones over all possible final states for the dominant partial wave $J_0$ in each final state:
\begin{equation}
 \sigma_{an} = \sum_i{\sigma_i^{(J_0)}}\ .
\end{equation}

Next, we can define the ``annihilation fraction'' $\kappa_i$ into the standard model particle pair $X_i - \bar{X}_i$:
\begin{equation}
\label{anfrac}
\kappa_i = \frac{\sigma_i^{(J_0)}}{\sigma_{an}}\ .
\end{equation}

By combining Eqs. (\ref{detbal}) and (\ref{anfrac}) we can obtain the following expression for the WIMP pair-production cross-section:
\begin{equation}
\label{forwback}
\sigma(X_i \bar{X}_i \rightarrow 2\chi) =
2^{2(J_0 - 1)}\kappa_i\sigma_{an}\frac{(2S_{\chi}+1)^2}{(2S_X+1)^2}\left(1 - \frac{4M_{\chi}^{2}}{s}\right)^{1/2+J_0}\ .
\end{equation}

\noindent
Now, a few remarks should be made about the validity of this formula:
\begin{itemize}
\item Equation (\ref{forwback}) is valid for WIMP pair-production taking place at
 center-of-mass energies just above the pair-production threshold.
\item The detailed balancing equation is valid if and only if the process under
 consideration is characterized by time-reversal and parity invariance. It is well
 known that weak interactions violate both of them, up to some degree, which we
 ignore in this treatment.

%\textbf{Note: this should be perhaps checked???}
\end{itemize}

A process of the form $X_i \bar{X}_i\longrightarrow \chi\chi$ is not visible in a
 collider, since WIMPs only manifest themselves as missing energy.
At least one detectable particle is required for the event to pass the
triggers.
An additional photon from initial state radiation (ISR) is required
to be recorded on tape: $X_i \bar{X}_i \longrightarrow \chi\chi\gamma$.
We can correlate the WIMP pair-production process to the radiative WIMP
pair-production for photons which are either soft or collinear with respect
to the colliding beams. In this case, the two processes are related
through \cite{Birkedal:2004xn}:
\begin{equation}
\label{correlation}
\frac{d\sigma(e^+ e^- \rightarrow 2\chi +\gamma)}{dx d\cos\theta} \approx {\cal{F}}(x,\cos\theta) \tilde{\sigma}(e^+e^-\rightarrow 2\chi)\ ,
\end{equation}
where $x = 2E_{\gamma}/\sqrt{s}$, $\theta$ is the angle between the photon direction and the
direction of the incoming electron beam, $\tilde{\sigma}$ is the WIMP pair-production cross-section produced at
the reduced center of mass energy $\tilde{s}=(1-x)s$, and $\cal{F}$ is defined as:
\begin{equation}
{\cal{F}}(x,\cos\theta)=\frac{\alpha}{\pi}\frac{1+(1-x)^2}{x}\frac{1}{\sin^2\theta}\ .
\end{equation}

\noindent
Now, by combining Eqs. (\ref{correlation}) and (\ref{forwback}) we get the master equation:
\begin{equation}
\label{ILCmastereq}
\frac{d\sigma}{dx d\cos\theta}(e^+e^- \rightarrow 2\chi+\gamma) \approx
\frac{\alpha\kappa_e \sigma_{an}}{16\pi} \frac{1+(1-x)^2}{x} \frac{1}{\sin^2\theta}2^{2J_0}(2S_\chi +1)^2
\left(1-\frac{4M_{\chi}^2}{(1-x)s}\right)^{1/2+J_0}\ .
\end{equation}

The problem is that very collinear photons fall outside the reach of any detector, due to practical
limitations in the coverage of the volume around the beam pipe. Also, typically, lower cuts are included in the detected
transverse momentum of photons, $p_T = E_\gamma\sin\theta$, in order to avoid excessive background signals at low energies.
So, if we are to use this approach, we have to examine its validity outside the soft/collinear region. The accuracy
of the collinear approximation for hard photons at all angles has been discussed in the original paper \cite{Birkedal:2004xn},
with the conclusion that the approach works quite well.

However,
an important point should be taken into account here. From the previous discussion on the validity of the method, we have
to impose specific kinematical cuts on the detected photons. We consider the following conditions:
\begin{itemize}
\item We demand an overall condition $\sin\theta \geq 0.1$ and $p_T \geq 7.5$ GeV in order to assure the detectability of the photons.
\item In order to assure the fact that any photon under examination corresponds to non-relativistic WIMPs,
we demand $v_{\chi}^2 \leq 1/2$. This gives a lower kinematical cut, along with an upper cut corresponding just to
the endpoint of the photon spectrum:
\begin{equation}
\label{cuts}
\frac{\sqrt{s}}{2}\left(1-\frac{8M_{\chi}^2}{s}\right) \leq E_{\gamma} \leq \frac{\sqrt{s}}{2}\left(1-\frac{4M_{\chi}^2}{s}\right)\ .
\end{equation}
\end{itemize}

These conditions present a flaw: the energy limits depend on the mass we wish to constrain. On the other hand,
for the reasons explained before, we cannot treat the signals without imposing such kinds of cuts, if we do not
want either to abuse the method or stick to heavy WIMPs (which, for kinematical reasons, cannot be relativistic).
The only way to evade this problem is to suppose that other
dark matter detection experiments (or, eventually, the LHC in the framework of specific models) will have already
provided us with some sort of limits on the WIMP mass. In this case, having an idea of the region in which the WIMP
mass falls, we can also estimate the cuts that will safely keep us outside the relativistic region and only consider
photons within this region.

The main source of background events is the standard model radiative neutrino production,
$e^+e^-\longrightarrow \nu\bar{\nu}\gamma$. Apart from these background events, various models predict
additional signals of the form ``$\gamma$ + missing energy'', one of the most well-known examples being radiative
sneutrino production \cite{Dreiner:2006sb, Dreiner:2007qc}, predicted in the framework of several supersymmetric models.
In the spirit of staying as model-independent as possible, we will ignore all possible beyond standard model processes.

\subsection{Basic Results}
\subsubsection{Non-polarized beams}
\begin{figure}
\begin{center}
\includegraphics[angle=270,scale=0.5]{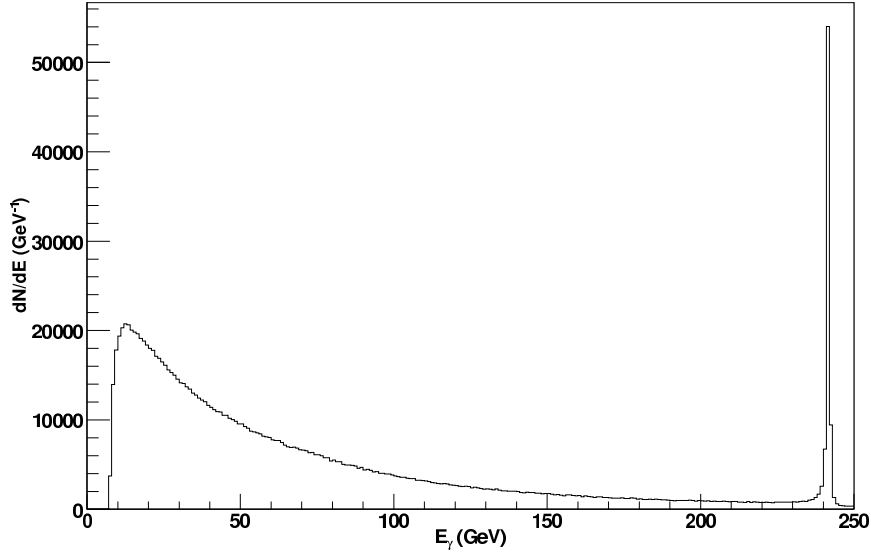}
\caption{{\footnotesize
Radiative neutrino production background $e^+e^-\rightarrow\nu\bar{\nu}\gamma$ for the ILC, for an unpolarized initial state.}}
\label{Fig.1}
\end{center}
\end{figure}

We place ourselves in the framework of the ILC
 project with a center-of-mass energy of $\sqrt{s} = 500$ GeV and an integrated
 luminosity of $500$ fb$^{-1}$. In order to estimate the background events, we
used the CalcHEP code \cite{Pukhov:1999gg, Pukhov:2004ca} to generate
$1.242.500$ $e^+e^-\longrightarrow \nu\bar{\nu}\gamma$ events, corresponding to
 the aforementioned conditions. The total radiative neutrino production
background can be seen in Fig.\ref{Fig.1}. The peak at
$E_\gamma=\sqrt{s}/2\cdot(1-M_Z^2/s)\simeq241.7$ GeV corresponds to the
radiative returns to the $Z$ resonance.

We generate a predicted ``observable'' spectrum for given values
of the WIMP mass and the annihilation fraction. During this study, we do not proceed to a (more realistic) full
detector simulation, as done for example in Ref. \cite{Bartels:2007cv}, but stick to prediction levels in order
to perform as thorough a scan as possible in the $(m_\chi,\kappa_e)$ parameter space and to have a picture of the
``a priori'' potential of the method.

\begin{figure}
\begin{center}
\includegraphics[angle=270,scale=0.5]{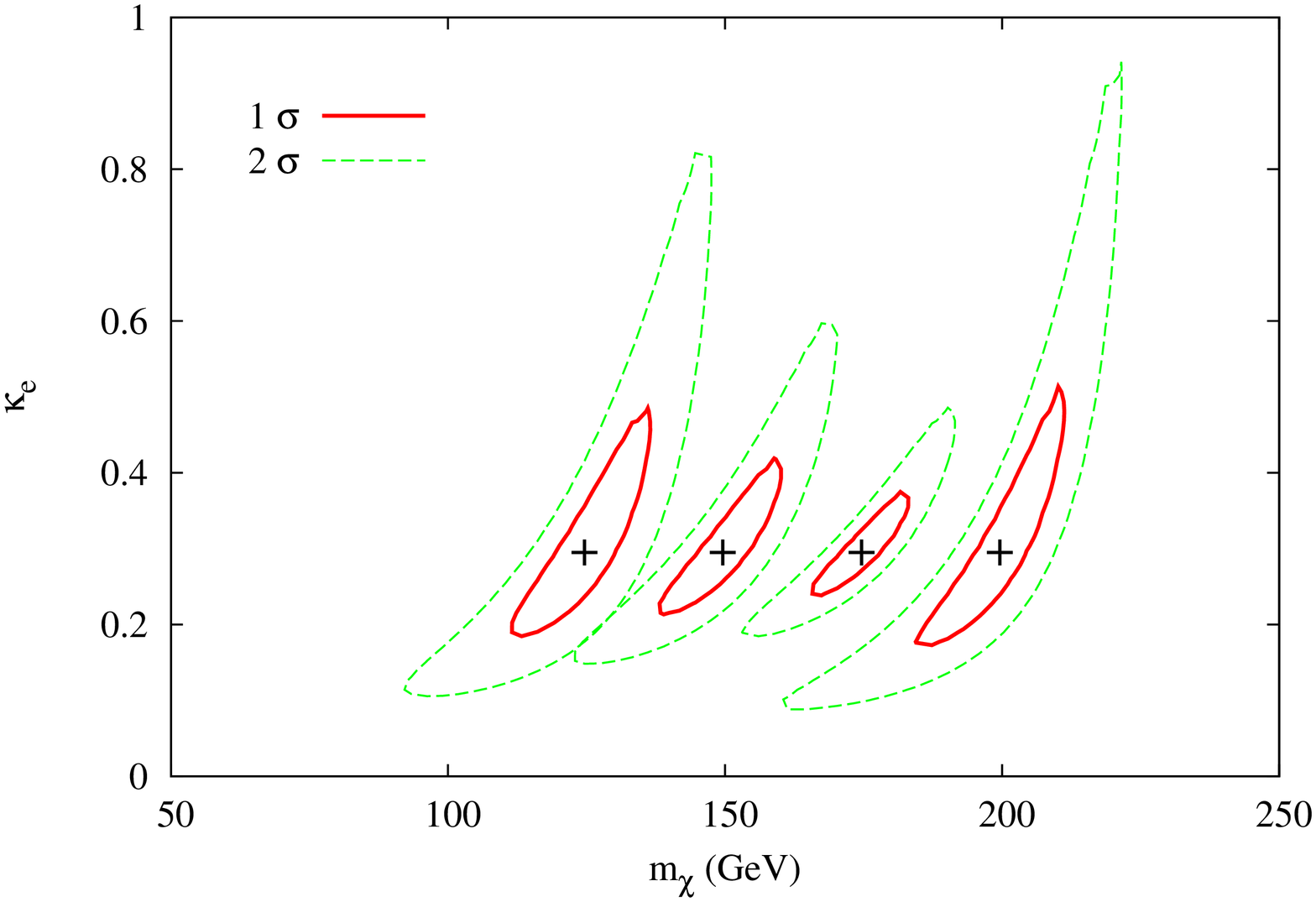}
\caption{{\footnotesize
Distribution of the maximum likelihood WIMP mass and annihilation fraction for the ILC in the  $(m_{\chi},\kappa_e)$ plane,
for $\kappa_e = 0.3$ and $m_\chi = 125, 150, 175$ and $200$ GeV. The inner (full lines) and outer (dashed lines) represent
the $68\%$  and $95\%$ CL region respectively.}}
\label{Fig.2}
\end{center}
\end{figure}

\begin{figure}
\begin{center}
\includegraphics[angle=270,scale=0.5]{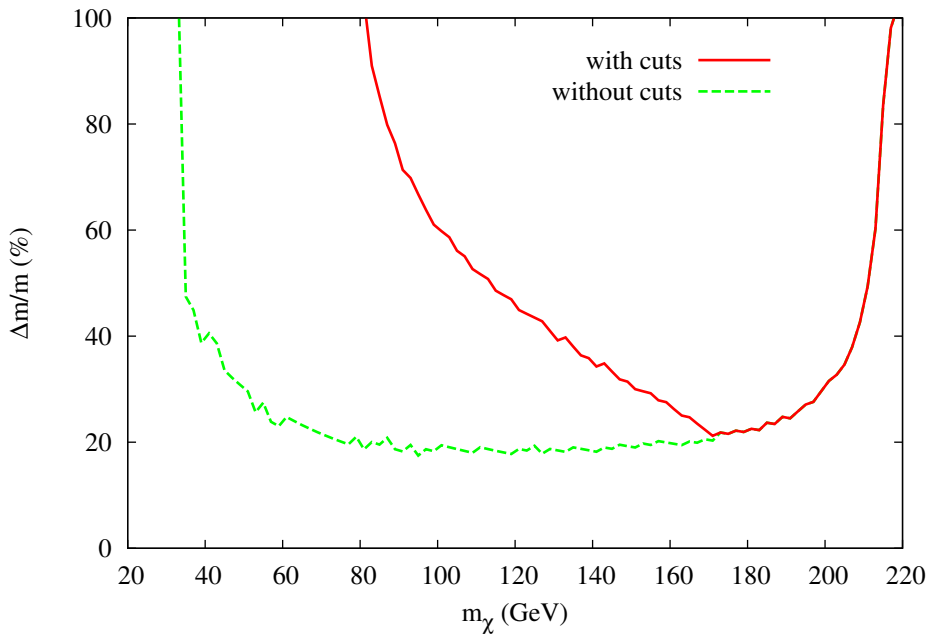}
\caption{{\footnotesize
Relative error in a generic WIMP mass determination, for $\kappa_e =0.3$ and at a %$2\sigma$
95\% confidence level.
The solid line corresponds to the results obtained after imposing the proper kinematical cuts, whereas the dashed
line to the case where we do not take these limits in consideration.}}
\label{Fig.3}
\end{center}
\end{figure}

Figure \ref{Fig.2} shows the predicted ability of the ILC to determine WIMP masses and annihilation
fractions for four points in the $(m_\chi, \kappa_e)$ parameter space for a $68\%$ and $95\%$ Confidence Level.
These results concern WIMPs with spin $S_\chi = 1/2$ and an angular momentum $J_0 = 1$ which corresponds to an
annihilation cross-section $\sigma_{an} = 7$ pb \cite{Birkedal:2004xn}. As can be seen, we are able to constrain quite
significantly the WIMP mass ($20\% - 40\%$ precision), while constraints on $\kappa_e$ are weaker.

Figure \ref{Fig.3} shows the relative error ($\Delta m_\chi/m_\chi$) for the mass
 reconstruction as a function of $m_\chi$, for $\kappa_e = 0.3$ and a $95\%$
 confidence level.  The solid line corresponds to the proper treatment including
 kinematical cuts. For indicative reasons, we also show the abused results obtained
 if we do not impose kinematical cuts on the photon energy (dashed line). The
 amelioration of the method's efficiency is obvious, although this is after all a
 false fact, since we include regions in which the approach is not valid. Above
$m_\chi \simeq 175$ GeV the two lines become identical, since the WIMPs cannot be
 relativistic. The capacity of the method peaks significantly for masses of the
 order of $175$ GeV because around this value we reach an optimal
combination of phase space volume and available spectrum that passes the kinematical cuts and can, hence,
be used for the calculation of the relevant $\chi^2$; whereas as we
move away from this value the accuracy tends to fall.

Let us make a final remark on the possibility of adopting a similar approach in the case of the LHC.
As we argued before, the large uncertainties entering the parton distribution functions and, hence, the large
uncertainty in the collision energy, affect significantly the precision of the whole procedure
(which is, already, based on approximations). Formally, in order to perform such a study for the LHC,
the computed cross-sections must be convoluted with the proton form factors. As an additional element,
the photon background in the LHC is expected to be much greater than in the ILC. The possibility of determining
WIMP properties through a model-independent method at the LHC has been addressed to in Ref. \cite{Feng:2005gj},
where the authors conclude that WIMP detection will be extremely difficult, if even possible.

\subsubsection{Polarized beams}
The reach of the ILC can be further increased by polarizing the beams.
For polarized beams, the signal cannot be fully characterized by $\kappa_e$; instead, four independent annihilation
fractions are needed, corresponding to the four possible $e^+e^-$ helicity configurations.

To apply Eq. (\ref{ILCmastereq}) to this case, we make the replacement:
\begin{eqnarray}
\kappa_e&\rightarrow&\frac{1}{4}(1+P_-)\,\left[(1+P_+)\,\kappa(e_-^Re_+^L)+(1-P_+)\,\kappa(e_-^Re_+^R)\right]\nonumber\\
&+&\frac{1}{4}(1-P_-)\,\left[(1+P_+)\,\kappa(e_-^Le_+^L)+(1-P_+)\,\kappa(e_-^Le_+^R)\right]\ ,
\end{eqnarray}
where $P_\pm$ are the polarizations of the positron and the electron beams.
As in ref \cite{Birkedal:2004xn,Bartels:2007cv}, let us assume that the WIMP couplings to electrons conserve both
helicity and parity: $\kappa(e_-^Re_+^L)=\kappa(e_-^Le_+^R)=2\,\kappa_e$ and $\kappa(e_-^Re_+^R)=\kappa(e_-^Le_+^L)=0$.

\begin{figure}
\begin{center}
\includegraphics[angle=270,scale=0.5]{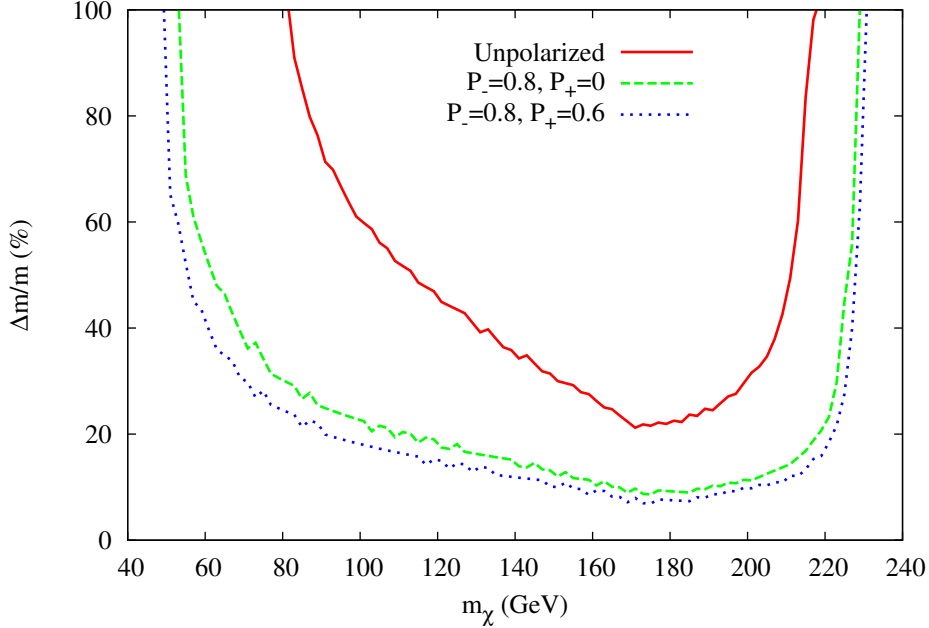}
\caption{{\footnotesize
Relative error in a generic WIMP mass determination, for three cases of beam polarization, including all proper kinematical cuts.}}
\label{Fig.4}
\end{center}
\end{figure}

In Fig.\ref{Fig.4} we show the relative error for the mass reconstruction for $\kappa_e=0.3$ and $95\%$ confidence level,
for the unpolarized scenario and for two different polarizations: $(P_-,P_+)=(0.8,0)$ and $(0.8,0.6)$.

%------------------------------------------------------------

%\noindent
%It is clear that there might be other model-independent methods that will provide
% better results both for the LHC and the ILC. In this paper, we addressed a simple
% case in which, through very simple formulae, we manage to get some interesting
% results on the possibility of constraining WIMP dark matter candidates.

\section{Complementarity}

\begin{figure}
\begin{center}
\includegraphics[angle=270,scale=0.5]{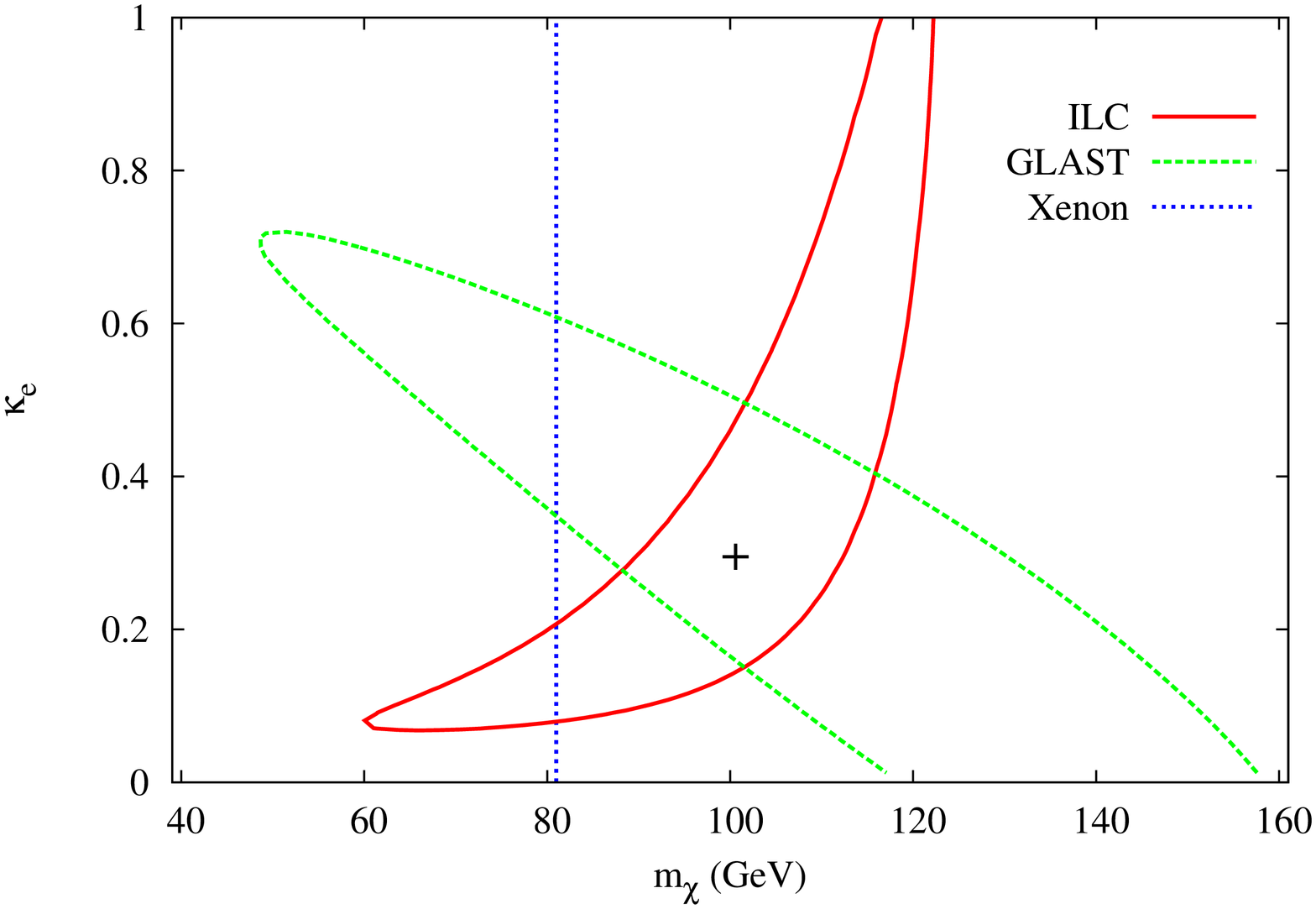}
\includegraphics[angle=270,scale=0.5]{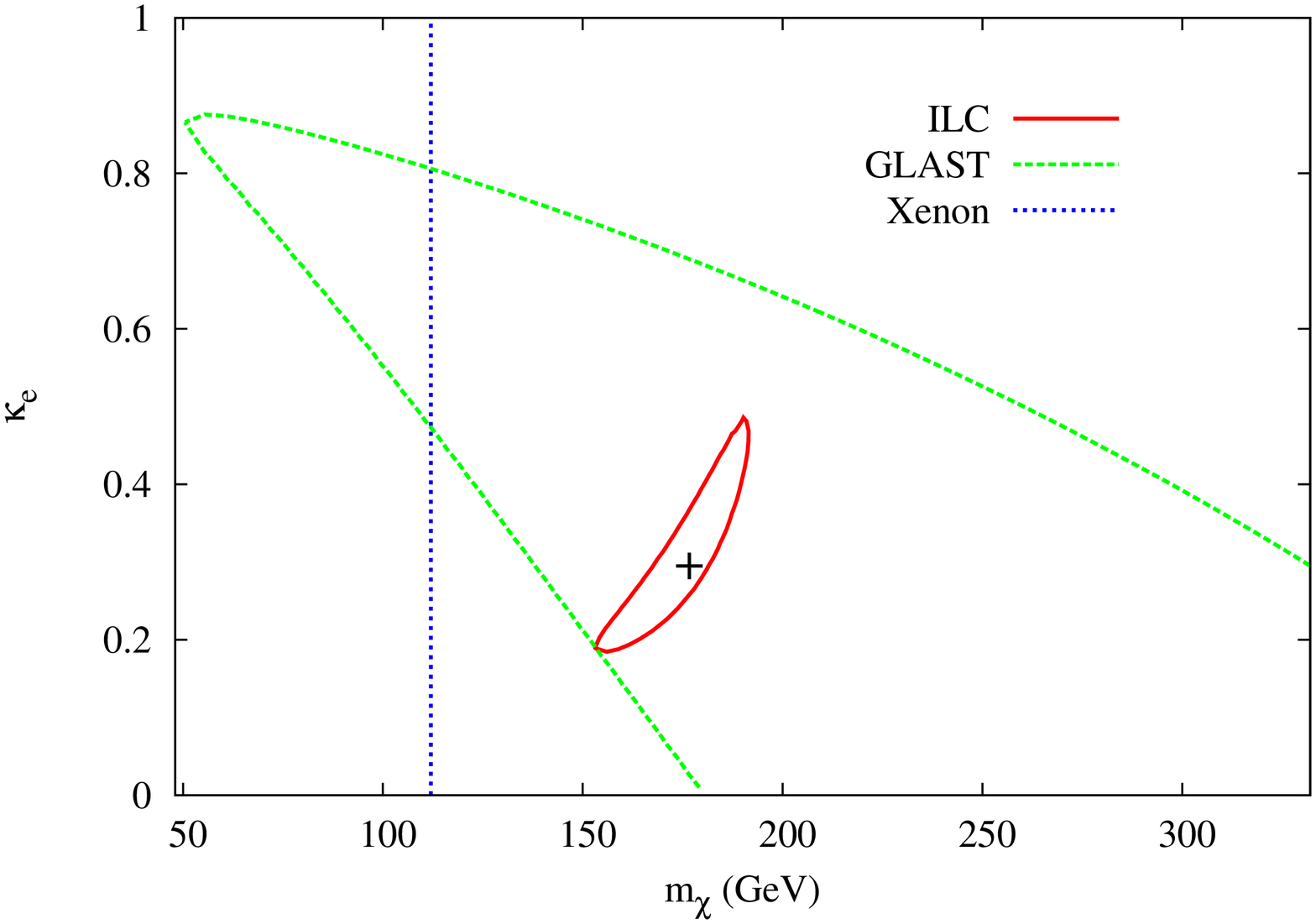}
\caption{{\footnotesize
Comparison between a $100$ kg XENON-like experiments (dotted line) with
$\sigma_{\chi-p}=10^{-8}$ pb, GLAST (dashed line) in the case of an NFW halo
 profile with $\langle\sigma v\rangle=3\cdot10^{-26}$ cm$^3$s$^{-1}$, and
 unpolarized ILC sensitivity (solid line) at $95\%$ of confidence level, for
 different WIMP masses $m_\chi=100$ and $175$ GeV, and $\kappa_e=0.3$.}}
\label{complementabananes}
\end{center}
\end{figure}

In Fig.\ref{complementabananes} we compare the precision levels for direct and
 indirect detection experiments, along with the corresponding results of the
method we followed for the ILC for two cases of WIMPs masses,
 $m_\chi = 100$ GeV and $175$ GeV, and $\kappa_e = 0.3$. We plot the
results in the $(m_\chi, \kappa_e)$ plane. This is done as the $\kappa_e$
parameter entering the ILC treatment presented before is, in fact,
the same parameter as the corresponding
branching ratio
$Br_i= \frac{\langle \sigma_i v \rangle}{\langle \sigma v \rangle}$
appearing in eq. (\ref{Eq:flux}) for $i=e$.
%($Br_e=1-\kappa_e$ to be precise)

\begin{center}
\begin{table}
\centering
\begin{tabular}{|c|c|c|c|}
\hline
$m_{\chi}$ & XENON & GLAST & ILC   \\
\hline
 $50$ GeV & $-5/+7$   GeV & $\pm 12$ GeV   & $-$ \\
$100$ GeV & $-19/+75$ GeV & $-50/+60$ GeV  & $-40/+20$ GeV\\
$175$ GeV & $-65/$    GeV & $-125$ GeV & $-20/+15$ GeV\\
$500$ GeV & $-$           & $-$            & $-$ \\
\hline
\end{tabular}
\caption{{\footnotesize Precision on a WIMP mass expected from the
different experiments at a $95\%$ CL after 3 years of exposure,
$\sigma_{\chi-p}=10^{-8}$ pb a NFW profile and a
500 GeV unpolarized linear collider with an integrated luminosity of 500$\mbox{fb}^{-1}$}}
\label{Tab:summary}
\end{table}
\end{center}

The blue-dotted line corresponds to a $100$ kg XENON-like experiment, where the
WIMP-nucleus cross-section has been assumed to be $10^{-8}$ pb. The
green-dashed line
 depicts the results for a GLAST-like experiment assuming a NFW halo profile.
The total annihilation cross-section into standard model particles has been
taken to be $\langle\sigma v\rangle=3\cdot10^{-26}$ cm$^3$s$^{-1}$. The red-plain
line represents our results for an ILC-like collider, with non-polarized beams.
All the results are plotted for a $95\%$ confidence level.

We can see that for different regions of the WIMP mass, the three kinds of
 experiments that we have
used as prototypes can act in a highly complementary way. For example, for
the case of a 100 GeV WIMP, indirect detection
or an ILC-like experiment alone can provide us with limited precision
both for the WIMP mass (of the order of $60\%$) and the $\kappa_e$ parameter
 (where the results are even worse). Combined measurements can dramatically
 increase the precision, reaching an accuracy of $25\%$ in mass. If we
additionally include direct detection measurement,
we can further increase the precision.

For the case of a $175$ GeV WIMP, a point where the unpolarized ILC sensitivity peaks, we see that the
dominant information comes from this source. Nevertheless, even if we only combine direct and indirect detection
experiments, we see that we can, in fact, acquire non-negligible constraints on the dark matter candidate mass.

To summarize the analysis, we show in Table \ref{Tab:summary} the precision expected for
several interesting dark matter masses.
Whereas a light WIMP (50 GeV) can be reached by both types of dark matter experiments
with a relatively high level of precision,
our analysis fails in the ILC case because of the relativistic nature
of the WIMP. On the contrary, the ILC would be particularly
efficient to discover and measure a WIMP with a mass of about 175 GeV.
Concerning a 500 GeV WIMP, which is kinematically unreachable at
the linear collider, it would be difficult to be observed by GLAST
or XENON. Only a lower bound could
be determined experimentally.

\section{Conclusions}

A Weakly Interacting Massive Particle (WIMP), with mass lying
from the GeV to the TeV scale, is one of the preferred candidates for the dark matter of the Universe.

We have discussed the possibility of identifying WIMP properties in a model-independent way. For that we have
considered direct and indirect searches, and in particular the interesting cases of a XENON-like 100 kg experiment and the GLAST satellite.
We have shown that
whereas direct detection experiments will probe efficiently
light WIMPs given a positive detection (at the $10\%$ level for
$m_\chi\lesssim 50$ GeV), GLAST will be able to confirm
and even increase the precision in the case of NFW profile, for a
WIMP-nucleon cross-section $\sigma_{\chi-p} \lesssim 10^{-8}$ pb.

Moreover, both XENON and GLAST are complementary with a future ILC project,
and the measurements from the three experiments will be able to
increase significantly the precision that we can reach on the mass of
the WIMP.

\vspace*{1cm}
\noindent{\bf Acknowledgements}

The authors want to thank particularly G. Bertone for the careful reading
of the work. N.B. and A.G. would also like to thank R. K. Singh and A. Djouadi
for useful discussions concerning the ILC part of the present work. Finally, A.G. would
like to thank Y. Amhis for useful discussions concerning the statistical treatment.

Likewise, the authors
would like to thank the ENTApP Network of the ILIAS project RII3-CT-2004-506222
and the French ANR project PHYS@COL\&COS for financial support.
Y.M. would like to thank the members of the
Institute for Theoretical Physics of Warsaw for their warm hospitality, and
financial support via the ``Marie Curie Host Fellowship for Transfer of
Knowledge'', MTKD-CT-2005-029466.
The work of A.G. is sponsored by the hepTOOLS Research Training Network
MRTN-CT-2006-035505.
The work of C.M. was supported
in part by the Spanish DGI of the
MEC
under Proyectos Nacionales FPA2006-01105 and FPA2006-05423,
by the European Union under the RTN programs
MRTN-CT-2004-503369 and UniverseNet MRTN-CT-2006-035863, and by
the Comunidad de Madrid under Proyecto HEPHACOS S-0505/ESP-0346.
The comments and suggestion of the referee were more than precious
for the evolutions of the work.

\newpage

\section*{Appendix}
In this Appendix we present the method followed in order to obtain the functions describing the standard
model particle decay into $\gamma$-rays.
In order to determine these spectral functions, we generated 300000 events of standard model particles decaying
(directly or through secondary decays) into $\gamma$-rays using the PYTHIA \cite{Sjostrand:2006za} package, taking
care in order to include all possible decay channels. Following the method of Ref. \cite{Hisano:2005ec}, we fitted the resulting spectra  through functions of the form:
\begin{equation}
\frac{dN_\gamma^i}{dx} = \exp\left[F_i(\,\ln(x)\,)\right],
\end{equation}
where $i$ represents the i-th WIMP annihilation channel, $i = WW, ZZ, \mbox{etc}$; $x = E_\gamma/m_\chi$ with $m_\chi$ being
the WIMP mass and $F$ are seventh-order polynomial functions which were found to be the following:
\begin{eqnarray*}
WW(x) & = & -7.72088528 -8.30185509\,x -3.28835893\,x^2 -1.12793422\,x^3\\
 & - & 0.266923457\,x^4 -0.0393805951\,x^5 -0.00324965152\,x^6 -0.000113626003\,x^7,\\
ZZ(x) & = & -7.67132139 -7.22257853\,x -2.0053556\,x^2 -0.446706623\,x^3 \\
 & - & 0.0674006343\,x^4 -0.00639245566\,x^5 -0.000372241746\,x^6 -1.08050617\cdot10^{-5}\,x^7,\\
b\bar{b}(x) & = & -11.4735403 -17.4537277\,x -11.5219269\,x^2 -5.1085887\,x^3\\
 & - & 1.36697042\,x^4 -0.211365134\,x^5 -0.0174275134\,x^6 -0.000594830839\,x^7,\\
u\bar{u}(x) & = & -4.56073856 -8.13061428\,x -4.98080492\,x^2 -2.23044157\,x^3\\
 & - & 0.619205713\,x^4 -0.100954451\,x^5 -0.00879980996\,x^6 -0.00031573695\,x^7,\\
d\bar{d}(x) & = & -4.77311611 -10.6317139\,x -8.33119583\,x^2 -4.35085535\,x^3\\
& - & 1.33376908\,x^4 -0.232659817\,x^5 -0.0213230457\,x^6 -0.000796017819\,x^7,\\
\tau^-\tau^+(x) & = & -5.64725113 -10.8949451\,x -7.84473181\,x^2 -3.50611639\,x^3\\
& - & 0.942047119\,x^4 -0.14691925\,x^5 -0.0122521566\,x^6 -0.000422848301\,x^7.
\end{eqnarray*}

\begin{figure}
\begin{center}
\includegraphics[angle=270,scale=0.5]{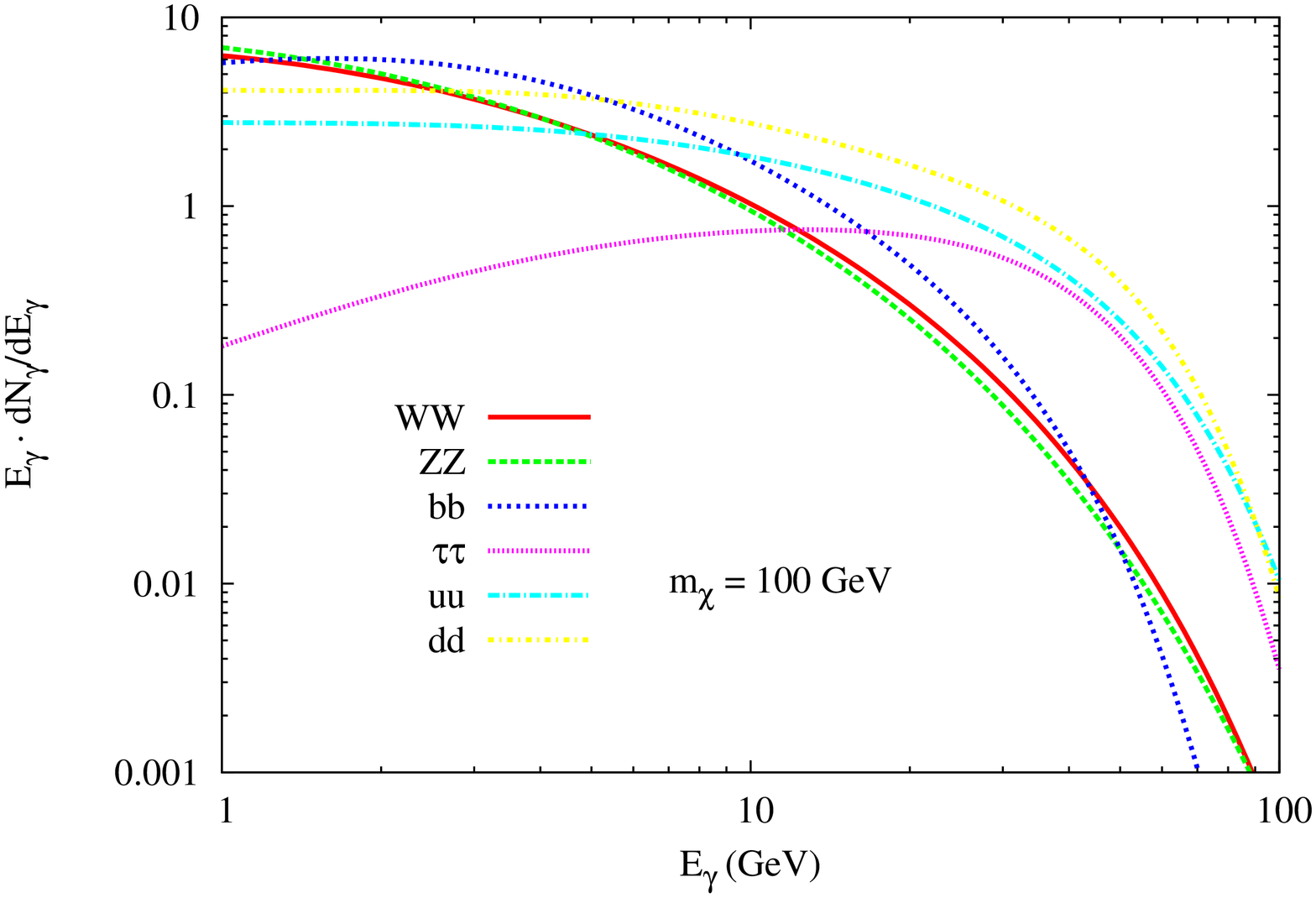}
\includegraphics[angle=270,scale=0.5]{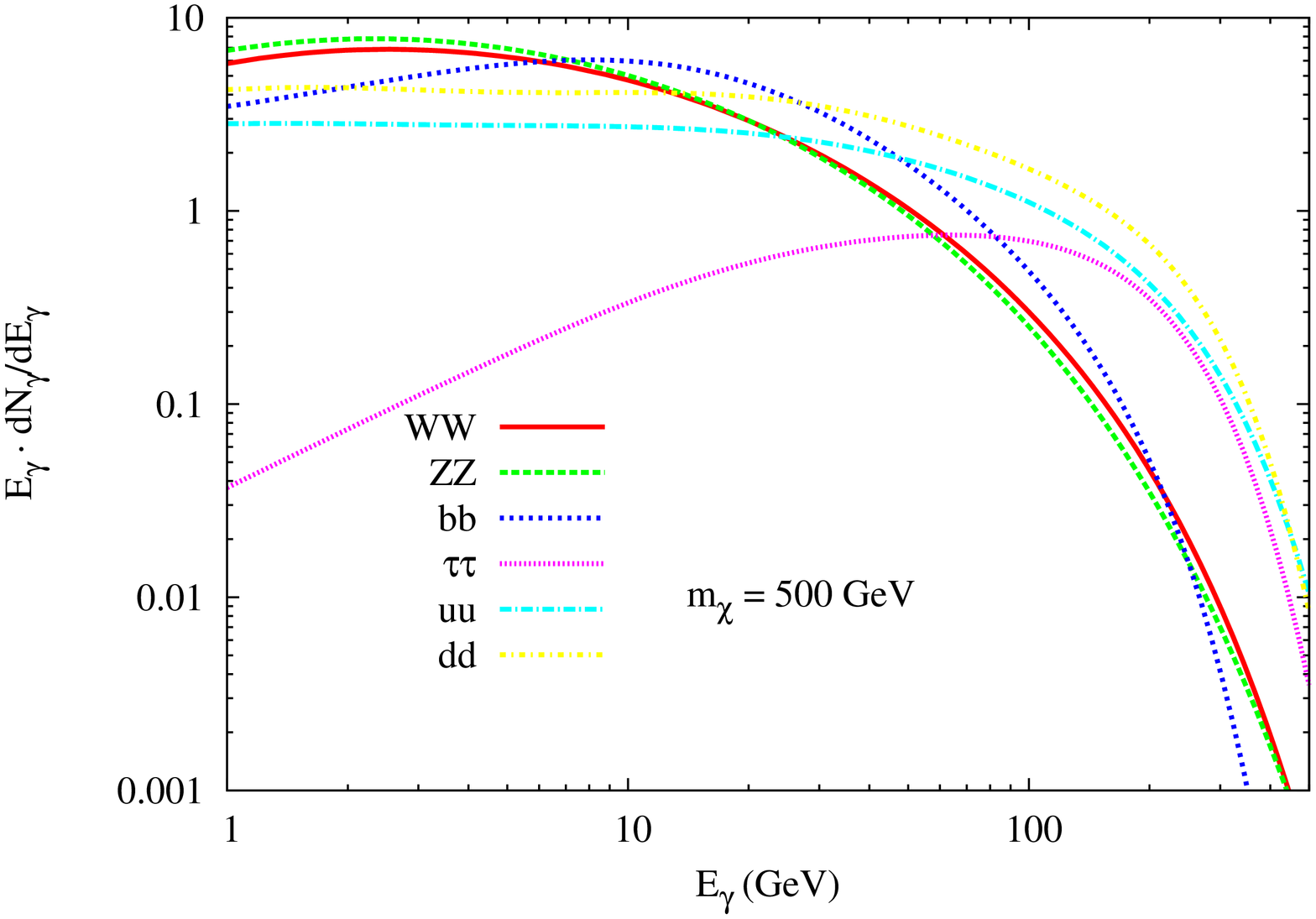}
\caption{{\footnotesize
Separate contributions from standard model particles decaying into $\gamma$-rays for $m_\chi=100$ and $500$ GeV.
The PYTHIA result points have been suppressed for
the sake of clarity.}}
\label{Fig.5}
\end{center}
\end{figure}

The case of WIMP annihilation into $\mu^+\mu^-$ pairs has a relatively small
decay contribution, to the photon spectrum,
coming from the $\mu\rightarrow e^-\bar{\nu}_e\nu_\mu\gamma$
channel, which has a small branching ratio. $e^+ e^-$ pair production
contributes to the gamma--ray spectrum through different (not decay) processes, mainly
inverse Compton scattering and synchrotron radiation. These contributions
depend crucially on the assumptions made concerning the intergalactic
medium and will not be analysed here (For a relevant treatment see, e.g.,
section V of \cite{Bertone:2004ps}).
This means, practically,
 that the $e^+e^-$ and $\mu^+\mu^-$ spectral functions are set equal to zero.
A graphical representation of these functions can be seen in Fig.\ref{Fig.5}.

These functions can afterward be used in order to generate any gamma-ray flux according to eq. (\ref{Eq:flux})

As we can see, all contributions are quite similar, apart from the $\tau^-\tau^+$ channel which has a
characteristic hard form. Nevertheless, at high energies, the form of all contributions becomes almost identical.

\newpage

\nocite{}
\bibliography{bmn}
\bibliographystyle{unsrt}

\end{document}